\documentclass[10pt,a4paper]{article}

\usepackage[margin=2cm]{geometry}
\usepackage{amsfonts}
\usepackage{graphicx}
\usepackage{subfig}
\usepackage{cite}
\usepackage{caption}
\usepackage{amsmath}
\usepackage{amssymb}
\usepackage{color}
\usepackage{authblk}
\usepackage[utf8]{inputenc}
\usepackage{epsfig}
\usepackage{float}
\title{\textbf{Kerr black holes with synchronised axionic hair}}
\author[1]{Jorge F. M. Delgado\footnote{jorgedelgado@ua.pt}}
\author[1]{Carlos A. R. Herdeiro\footnote{herdeiro@ua.pt}} \author[1]{Eugen Radu\footnote{eugen.radu@ua.pt}}
\affil[1]{\normalsize Departamento de Matemática da Universidade de Aveiro and 

Center for Research and Development in Mathematics and Applications -- CIDMA

Campus de Santiago, 3810-183 Aveiro, Portugal}

\begin{document}

\date{December 2020}

\maketitle

\begin{abstract}
\normalsize

We construct and analyse Kerr black holes (BHs) with synchronised axionic hair. These are the BH generalisations of the recently constructed rotating axion boson stars \cite{Delgado:2020udb}. Such BHs are stationary, axially symmetric, asymptotically flat solutions of the complex Einstein-Klein-Gordon theory with a QCD axion-like potential. They are regular everywhere on and outside the event horizon. The potential is characterised by two parameters: the mass of the axion-like particle, $m_a$ and the decay constant $f_a$. The limit $f_a \rightarrow \infty$ recovers the original example of Kerr BHs with synchronised scalar hair~\cite{Herdeiro:2014goa}. The effects of the non-linearities in the potential become important for $f_a \lesssim 1$. We present an overview of the parameter space of the solutions together with a study of their basic geometric and phenomenological properties, for an illustrative value of the coupling that yields a non-negligible impact of the self-interactions.

\end{abstract}

\section{Introduction}

Testing the Kerr hypothesis is a central goal of current strong gravity research~\cite{Berti:2015itd,Barack:2018yly}. This is the hypothesis that astrophysical black holes  (BHs), when near equilibrium, are well described by the Kerr metric \cite{Kerr:1963ud}. This working assumption is intimately connected with the no-hair conjecture \cite{Ruffini:1971bza}, which states that the dynamically formed equilibrium BHs have no other macroscopic degrees of freedom beyond those associated with Gauss laws (and hence gauge symmetries) -- see $e.g.$ \cite{Herdeiro:2015waa,Sotiriou:2015pka,Volkov:2016ehx,Cardoso:2016ryw} for reviews. 

Over  the  last  few  years it  has been realized  that the Kerr hypothesis can be challenged  \textit{even} within  General  Relativity (GR) and  \textit{even}  with physically simple and reasonable energy-matter contents, due to the discovery that free, massive complex scalar or vector fields can endow the Kerr BH with  synchronised  bosonic ``hair"~\cite{Herdeiro:2014goa,Herdeiro:2015gia,Herdeiro:2016tmi,Santos:2020pmh}, see, $e.g.$~\cite{Herdeiro:2015tia,Delgado:2016jxq,Herdeiro:2018daq,Wang:2018xhw,Herdeiro:2018djx,Kunz:2019bhm,Kunz:2019sgn,Delgado:2019prc,Collodel:2020gyp} for generalizations. Moreover, if these fields are sufficiently ultra-light, this ``hair" could occur in the mass range of astrophysical BH candidates, spanning the interval from a few solar masses, $\sim M_\odot$ (stellar mass BHs),  to $\sim 10^{10} \ M_\odot$  (supermassive BHs).

The existence of such hairy BHs circumvents various no-scalar (and no-Proca) hair theorems - see $e.g.$~\cite{Bekenstein:1972ny,Bekenstein:1996pn,Herdeiro:2015waa}. Notwithstanding, the key test to the no-hair conjecture is a dynamical one. Are these hairy BHs dynamically robust? That is, can they form dynamically and be sufficiently stable? The current understanding is that: 1) there are dynamical formation channels, namely: $a)$ the superrading instability~\cite{Brito:2015oca} of the Kerr solution~\cite{East:2017ovw,Herdeiro:2017phl} and $b)$  mergers of bosonic stars~\cite{Sanchis-Gual:2020mzb}; 2) these hairy BHs may, themselves, still be afflicted by superradiant instabilities~\cite{Herdeiro:2014jaa,Ganchev:2017uuo} but these can be very long-lived, with a time scale that can exceed a Hubble time~\cite{Degollado:2018ypf}. The evidence so far, therefore, indicates BHs with synchronised ultralight bosonic hair are an interesting challenge to the Kerr hypothesis even in GR. This has led to different  phenomenological studies, including the study of their shadows \cite{Cunha:2015yba,Cunha:2016bpi,Cunha:2019ikd} and their $X$-ray phenomenology, namely the iron K$\alpha$ line \cite{Ni:2016rhz,Zhou:2017glv} and QPOs~\cite{Franchini:2016yvq}.

An important outstanding issue is the fundamental physics nature of the putative ultralight bosonic field that could endow BHs with ``hair". Possible embeddings in high energy physics have been suggested, such as the string axiverse~\cite{Arvanitaki:2009fg}, which proposes a landscape of ultralight axion-like particles emerging from string compactifications. If these particles have sufficiently small couplings with standard model particles, they are dark matter, only observable via their gravitational effects, one of which would be endowing spinning BHs  with synchronised hair. The QCD axion, in which these axion-like particles are inspired, is a pseudo Nambu-Goldstone boson suggested by the Peccei-Quinn mechanism \cite{Peccei:1977hh} to solve the strong CP problem in QCD \cite{tHooft:1976snw,Jackiw:1976pf}. It possesses a self-interactions  potential~\cite{Weinberg:1977ma,Wilczek:1977pj},  characterise by two parameters: the mass of the scalar field $m_a$ and the decay constant $f_a$. Thus, it would be interesting to assess the gravitational effects of such a  potential   for the axion-like particles that could endow spinning BHs with synchronised hair. This is the goal of the present paper.

Kerr BHs  with synchronised bosonic hair have a solitonic limit, obtained when taking  the BH  horizon to zero size,  wherein they reduce  to spinning boson  stars.  In the  case of the  original model  with a free, complex scalar field, the solitonic limit yields the so-called, spinning \textit{mini} boson stars~\cite{Schunck:1996he,Yoshida:1997qf,Schunck:2003kk}. Recently, axion boson  stars have been constructed, wherein  the complex scalar field is under the action of a QCD axion-like potential~\cite{Guerra:2019srj,Delgado:2020udb}, with the two aforementioned parameters.\footnote{Axionic stars with real fields have  been  considered,  $e.g.$ in~\cite{Hertzberg:2018lmt,Visinelli:2017ooc}.}   In the limit when $f_a \rightarrow \infty$, the potential reduces to a simple mass term, and the axion boson stars reduce to mini boson stars. Thus, in the same way, that the hairy BHs in~\cite{Herdeiro:2014goa} are the BH generalisation of mini boson stars, we shall construct here the BH generalisations of the rotating axion boson stars in~\cite{Delgado:2020udb}. Moreover, we shall study the basic physical properties and phenomenology of these BHs, hereafter dubbed \textit{Kerr BHs with synchronised axionic hair} (KBHsAH), or simply ``axionic BHs".

This work is organised as follows. In Section 2 we introduce the model together with the equations of motion and the \textit{ansatz} we used to solve them; we also present the QCD potential. In Section 3 we display the numerical framework  to tackle the field equations and to obtain the BH solutions, discuss the boundary conditions of the problem and how to extract physical quantities from the data. In Section 4 we show the numerical results, presenting the domain of existence of the numerical solutions together with an analysis of some of their physical properties and phenomenology. In Section 5 we present conclusions and some final remarks.

\section{The model}

The theory for which we shall obtain hairy BH solutions is the Einstein-Klein-Gordon model, that describes a massive complex scalar field, $\Psi$, minimally coupled to Einstein's gravity. The action of the theory is, using units such that $G = c = \hbar = 1$,
\begin{equation}\label{Eq:Action}
	\mathcal{S} = \int d^4 x \sqrt{-g} \left[ \frac{R}{16\pi} - g^{\mu\nu} \partial_{\mu} \Psi^* \partial_{\nu} \Psi - V(|\Psi|^2) \right] \ ,
\end{equation}
where $R$ is the Ricci scalar, $\Psi$ is the complex scalar field (`*' denotes complex conjugation) and $V$ is the scalar self-interaction potential. The equations of motion resulting from the variation of the action with respect to the metric, $g_{\mu\nu}$, and scalar field, are,
\begin{eqnarray}
	&E_{\mu\nu} \equiv R_{\mu\nu} - \frac{1}{2} g_{\mu\nu} R - 8\pi \ T_{\mu\nu} = 0 \ ,& \label{Eq:FieldEquationsMetric} \\
	&\Box \Psi - \dfrac{d V}{d |\Psi|^2} \Psi = 0 \ ,& \label{Eq:FieldEquationsScalarField}
\end{eqnarray}
where 
\begin{equation}
	T_{\mu\nu} = 2 \partial_{(\mu} \Psi^* \partial_{\nu)} \Psi - g_{\mu\nu} \left( \partial^\alpha \Psi^* \partial_\alpha \Psi + V\right) \ ,
\end{equation}
is the energy-momentum tensor associated with the scalar field.

A global $U(1)$ transformation $\Psi \rightarrow e^{i\chi} \Psi$, where $\chi$ is a constant, leaves the above action invariant; thus it is possible to write a scalar 4-current \cite{Herdeiro:2014goa}, $j^\mu = -i \left( \Psi^* \partial^\mu \Psi - \Psi \partial^\mu \Psi^* \right)$ which is conserved:  $D_\mu j^\mu = 0$. The existence of this symmetry and conserved current implies the existence of a conserved quantity -- the \textit{Noether charge} -- that can be computed by integrating the timelike component of the 4-current,
\begin{equation}
	Q = \int_\Sigma j^t \ .
\end{equation}
This quantity is interpreted as the number of scalar particles in a given solution, albeit this relation only becomes rigorous after field quantisation. For solitonic solutions (without an event horizon), moreover, $Q$ is related with the total angular momentum as~\cite{Yoshida:1997qf,Schunck:1996he} 
\begin{equation}
	J = m Q \ .
\end{equation}
This is a generic relation for rotating boson stars, already observed in other models with a self-interactions potential - see, $e.g.$~\cite{Kleihaus:2007vk}.

As BH generalisations of the work done on rotating axion boson stars~\cite{Delgado:2020udb}, we are interested in stationary, regular on and outside the event horizon, axi-symmetric and asymptotically flat solutions of the above equations of motion. The spacetime generated by these solutions possesses two commuting Killing vector fields, $\xi$ and $\eta$, which, in a suitable coordinate system, can be written as $\xi = \partial_t$ and $\eta = \partial_\varphi$ corresponding to stationarity and axi-symmetry, respectively. With these Killing vector fields, we can define the following \textit{ansatz} for the metric and the scalar field, written in terms of the spheroidal coordinates, $\{t, r, \theta, \varphi\}$,
\begin{eqnarray}
	&ds^2 = -e^{2F_0} N dt^2 + e^{2F_1} \left( \frac{dr^2}{N} + r^2 d\theta^2 \right) + e^{2F_2} r^2 \sin^2 \theta \left( d\varphi - W dt \right)^2 \hspace{2pt} , \hspace{10pt} N \equiv 1 - \frac{r_H}{r} \hspace{2pt} , \label{Eq:AnsatzMetric} \\
	&\Psi = \phi\ e^{i\left(\omega t - m \varphi \right)}  \label{Eq:AnsatzScalarField}~,
\end{eqnarray}
where $ \{F_1, F_2, F_0, W; \phi \}$ are \textit{ansatz} functions that depend exclusively on the $(r,\theta)$ coordinates; $r_H$ is the radial coordinate of the event horizon; $\omega$ is the angular frequency of the scalar field; and $m = \pm 1, \pm 2, \dots$ is the azimuthal harmonic index of the scalar field.

In this work, we aim to study BHs with a surrounding axion-like scalar field; thus we specify the scalar self-interaction potential as describing axion interactions. Following~\cite{Guerra:2019srj, Delgado:2020udb}, we will use the QCD axion potential~\cite{diCortona:2015ldu} to describe the axion interactions, added to a constant term in order to obtain asymptotically flat solutions. The potential in question has the form,
\begin{equation}\label{Eq:AxionPotential}
	V(\phi) = \frac{2 \mu_a^2 f_a^2}{B} \left[ 1 - \sqrt{1 - 4 B \sin^2 \left( \frac{\phi}{2 f_a} \right)} \right]~,
\end{equation}
where $B = \frac{z}{(1+z)^2} \approx 0.22$ is a constant, in which $z \equiv m_u/m_d \approx 0.48$ is the ratio between the up and down masses, $m_u$ and $m_d$, respectively; $\mu_a$ and $f_a$ define the axion-like particle (ALP) mass and quartic self-interaction coupling, respectively, and are two free parameters. This can be understood by performing an expansion of the potential around $\phi = 0$,
\begin{equation}\label{Eq:AxionPotentialExpansion}
	V(\phi) = \mu_a^2 \phi^2 - \left( \frac{3B-1}{12} \right) \frac{\mu_a^2}{f_a^2} \phi^4 + \frac{1+15 B(3B-1)}{360 f_a^4}\mu_a^2 \phi^6+\dots ~.
\end{equation}  
In this way, we confirm  $\mu_a$ is the ALP mass and $f_a$ is the quartic self-interaction coupling,
\begin{equation}
	m_a = \mu_a \hspace{2pt} , \hspace{10pt} \lambda_a = - \left( \frac{3B - 1}{12} \right) \frac{\mu_a^2}{f_a^2} \hspace{2pt}.
\end{equation}
In this work, we shall refer to $\mu_a$ and $f_a$ as the ALP mass and decay constant, respectively. 
We note that the above expansion is only valid on the regime where $ \phi \ll f_a $.
 In fact, to leading order, only the mass term remains. Therefore, as already noted in \cite{Delgado:2020udb} for the case of boson stars, for sufficiently large decay constant $f_a$, we expect the BHs solutions to become very similar to the original Kerr BHs with synchronised scalar hair, obtained in \cite{Herdeiro:2014goa}.

In a similar fashion to the family of Kerr BHs with synchronised scalar hair, the solutions obtained in the work are  possible due to the so-called \textit{synchronisation condition}. This condition can be interpreted as a synchronisation between the angular velocity of the event horizon, $\Omega_H$, corresponding to the value of the metric function $W$ at the horizon, $\Omega_H=W(r_H)$, and the phase angular velocity of the scalar field, $\omega/m$, $cf.$~\eqref{Eq:AnsatzScalarField}:
\begin{equation}
\label{conds}
	\omega = m \Omega_H \hspace{2pt} .
\end{equation}

Finally, let us mention that, as remarked in 
\cite{Delgado:2020udb},
the potential 
(\ref{Eq:AxionPotential})
allows
for the existence
of solutions even in the absence of the gravity term in the action (\ref{Eq:Action}).
The simplest case corresponds to (non-gravitating) Q-ball-like solitons in a  flat spacetime background.
As expected, these solutions possess generalizations on a Kerr BH background.
These bound states are in synchronous rotation with the BH horizon, $i.e.$ they still obey the  condition
(\ref{conds}) and share most of the properties of the non-linear Q-clouds in \cite{Herdeiro:2014pka}.
A discussion of these aspects will be
reported elsewhere.


\section{Framework}

\subsection{Boundary conditions}

To obtain numerical BH solutions appropriate boundary conditions must be imposed that enforce the sough physical behaviours. Such boundary conditions will now be summarised.
\begin{itemize}
	\item \textit{Asymptotically boundary conditions:} Asymptotic flatness implies that all \textit{ansatz} functions must go to zero asymptotically,
		\begin{equation}
			\lim_{r \rightarrow \infty} F_i = \lim_{r \rightarrow \infty} W = \lim_{r \rightarrow \infty} \phi = 0  \hspace{2pt} .
		\end{equation}	
		
	\item \textit{Axial boundary conditions:} Axial symmetry, together with regularity on the symmetry axis, implies that,
		\begin{equation}
			\partial_\theta F_i = \partial_\theta W = \partial_\theta \phi = 0 \hspace{2pt}, \hspace{10pt} \text{at} \hspace{10pt} \theta = \{0, \pi \}  \hspace{2pt} .
		\end{equation}
		
		Furthermore, we require the absence of conical singularities, thus, $F_1 = F_2$ on the axis. Since we focus on even parity solutions, which typically correspond to the fundamental solutions, they are symmetric \textit{w.r.t.} the equatorial plane. Hence, one only needs to solve the equations of motion in the range $0 \leqslant \theta \leqslant \pi/2$ and impose the following boundary conditions,
		\begin{equation}
			\partial_\theta F_i = \partial_\theta W = \partial_\theta \phi = 0 \hspace{2pt} , \hspace{10pt} \text{at} \hspace{10pt} \theta = \frac{\pi}{2}  \hspace{2pt} .
		\end{equation}
	
	\item \textit{Event horizon boundary conditions:} To simplify the study of these boundary conditions, let us introduce a radial coordinate transformation, $x = \sqrt{r^2 - r_H^2}$. With this coordinate transformation, we can perform a series expansion of the \textit{ansatz} functions at the horizon, $x = 0$, and find that,
		\begin{eqnarray}
			&&F_i = F_i^{(0)} + x^2 F_i^{(2)} + \mathcal{O}(x^4)  \hspace{2pt} , \\
			&&W = \Omega_H + \mathcal{O}(x^2)  \hspace{2pt} , \\
			&&\phi = \phi^{(0)} + \mathcal{O}(x^2)  \hspace{2pt} ,
		\end{eqnarray}
		where  $F_i^{(0)}, F_i^{(2)},\phi^{(0)}$ are functions of $\theta$.
		With these series expansions, we can naturally impose the following boundary conditions,
		\begin{equation}
			\partial_x F_i = \partial_x \phi = 0 \hspace{2pt} , \hspace{5pt} W = \Omega_H \hspace{2pt} , \hspace{10pt} \text{at} \hspace{10pt} r = r_H
		\end{equation}
\end{itemize}

\subsection{Extracting physical quantities}

The main physical quantities of interest for our analysis are encoded in the metric functions evaluated either at the horizon or at spacial infinity. For the former, we obtain the horizon angular velocity by using the horizon boundary condition mentioned in the previous sections, $\Omega_H = W|_{r_H}$, together with Hawking temperature, $T_H$, as well as the horizon area, $A_H$, through the following expressions,
\begin{equation}
	T_H = \frac{1}{4\pi r_H} e^{(F_0 - F_1)|_{r_H}} \hspace{2pt} , \hspace{10pt} A_H = 2\pi r_H^2 \int_0^\pi d\theta\ \sin \theta\ e^{(F_1 + F_2)|_{r_H}} \ .
\end{equation}
The entropy of the computed BH follows from the Bekenstein-Hawking formula, $S = A_H/4$.  
For the latter, we can compute the ADM mass, $M$, and angular momentum, $J$, through the asymptotic behaviour of $g_{tt}$ and $g_{t\varphi}$,
\begin{equation}
	g_{tt} = -e^{2F_0} N + e^{2F_2} W^2 r^2 \sin^2 \theta \rightarrow -1 + \frac{2M}{r} + \dots \hspace{2pt}, \hspace{10pt} g_{t\varphi} = -e^{2F_2} W r^2 \sin^2 \theta \rightarrow - \frac{2J}{r} \sin^2 \theta + \dots
\end{equation}

The above quantities are mutually related  through a Smarr-type formula \cite{Smarr:1972kt},
\begin{equation}\label{Eq:SmarrRelation}
	M = 2 T_H S + 2 \Omega_H \left( J - J^\Psi \right) + M^\Psi \hspace{2pt},
\end{equation}
where we have introduced two new quantities, which can not be computed directly from the horizon or spacial infinity data,
\begin{equation}
	M^\Psi = -2 \int_\Sigma dS_\mu \left( T^\mu_\nu \xi^\nu - \frac{1}{2} T \xi^\mu \right) \hspace{2pt} , \hspace{10pt}  J^\Psi = \int_\Sigma d S_\mu \left( T^\mu_\nu \eta - \frac{1}{2} T \eta^\mu \right) \hspace{2pt} ,
\end{equation}
corresponding to the scalar field mass and angular momentum, respectively\footnote{The scalar field potential $V(|\Psi|^2)$ enters (\ref{Eq:SmarrRelation}) via the  $M^\Psi$-term.}. 
In their definition, $\Sigma$ is a spacelike surface, bounded by the 2-sphere at infinity, $S^2_\infty$, and the spatial section of the horizon, $H$. Moreover, the angular momentum of the scalar field is related to the Noether charge which arises from the global $U(1)$ symmetry of the scalar field, $Q$, as $J^\Psi = m Q$. For a solution composed entirely of axionic scalar hair (no horizon), the ADM angular momentum must be equal to the angular momentum of the scalar field; likewise, for a solution without axionic scalar hair, the ADM angular momentum of the solution equals the horizon angular momentum. Thus, in order to evaluate how hairy a given BH solution is, we define the following dimensionless parameter:
\begin{equation}
	q \equiv \frac{J^\Psi}{J} = \frac{m Q}{J} \hspace{2pt}.
\end{equation}
One easily sees that, when $q = 0$, we have a bald BH, corresponding to a Kerr BH. On the other end, when $q=1$, we have a solution entirely compose of axionic hair, which corresponds to a rotation axion boson star \cite{Delgado:2020udb}. For any other value $0 < q < 1$, we have a rotating BH surrounded by a non-trivial, backreacting, massive rotating axion scalar field.

\subsection{Numerical approach}

To perform the numerical integration of the equations of motion resulting from Eqs. \eqref{Eq:FieldEquationsMetric} and \eqref{Eq:FieldEquationsScalarField} with the \textit{ansatz} Eqs. \eqref{Eq:AnsatzMetric} and \eqref{Eq:AnsatzScalarField}, it is useful to rescale key quantities by $\mu_a$,
\begin{equation}
	r \rightarrow r \mu_a \hspace{2pt}, \hspace{10pt} \phi \rightarrow \phi \sqrt{4\pi} \hspace{2pt} , \hspace{10pt} \omega \rightarrow \omega/\mu_a~,
\end{equation}
together with $ f_a \rightarrow f_a \sqrt{4\pi}$.
This leads to the disappearance of the ALP mass constant
from the equations of motion numerically solved, but all global quantities will be express in terms of $\mu_a$.

In our approach, by expanding the equations of motion, we get a set of five coupled, non-linear, elliptic partial differential equations for the \textit{ansatz} functions, $\mathcal{F}_a = \{ F_0, F_1, F_2, W; \phi \}$. They are compose of the Klein-Gordon equation, Eq. \eqref{Eq:FieldEquationsScalarField}, together with the following combination of the Einstein equations, Eq. \eqref{Eq:FieldEquationsMetric},
\begin{eqnarray}
	&&E^r_r + E^\theta_\theta - E^\varphi_\varphi - E^t_t = 0 \hspace{2pt} , \\
	&&E^r_r + E^\theta_\theta - E^\varphi_\varphi + E^t_t + 2W E^t_\varphi = 0 \hspace{2pt} , \\
	&&E^r_r + E^\theta_\theta + E^\varphi_\varphi - E^t_t - 2W E^t_\varphi = 0 \hspace{2pt} , \\
	&&E^t_\varphi = 0 \hspace{2pt} .
\end{eqnarray}
The remaining equations, $E^r_\theta = 0$ and $E^r_r - E^\theta_\theta = 0$ are not solved directly, instead, they are used as constraint equations to evaluate the accuracy of the numerical solution. Typically they are satisfied at the level of the overall numerical accuracy. 

Our numerical treatment can be summarised as follows. We restrict the domain of integration to the region outside the horizon. Using the aforementioned radial coordinate transformation $x = \sqrt{r^2 - r_H^2}$, we transform the radial region of integration from $[r_H, \infty)$ to $[0,\infty)$. Then, we introduce a new radial coordinate that maps the semi-infinite region $[0,\infty)$ to the finite region $[0,1]$. Such map can be defined in several ways, but in this work we choose the new coordinate, $\bar{x} = x/(x+1)$. After this, the equations $\mathcal{F}_a$ are discretised on a grid in $\bar{x}$ and $\theta$. Most of the results presented here were obtained on an equidistant grid with  $251 \times 30$ points. The grid covers the integration region $0 \leqslant \bar{x} \leq 1$ and $0 \leqslant \theta \leq \pi/2$.

The equations of motion have been solved subject to the boundary conditions introduced above by using a professional package, entitled \textsf{FIDISOL/CADSOL}~\cite{schoen}, which employs a Newton-Raphson method with an arbitrary grid and consistency order. This code uses the finite difference method, providing also an error
estimate for each unknown function. For the solutions in this work, the maximal numerical error for the functions is estimated to be on the order of $10^{-3}$. The Smarr relation, Eq. \eqref{Eq:SmarrRelation}, provides a further test of the numerical accuracy, leading to error estimates of the same order.

In our scheme, there are four input parameters: \textbf{i)} the decay constant $f_a$ in the potential, Eq. \eqref{Eq:AxionPotential}; \textbf{ii)} the angular frequency of the scalar field $\omega$; \textbf{iii)} the azimuthal harmonic index $m$; and \textbf{iv)} the radial coordinate of the event horizon $r_H$. The number of nodes $n$ of the scalar field, as well as all other quantities of interest mentioned before, are computed from the numerical solution. For simplicity, we have restricted our study to the fundamental configurations, \textit{i.e.} with a nodeless scalar field, $n = 0$ and with $m = 1$. Also, from the results presented in \cite{Delgado:2020udb}, we shall illustrate the effect\footnote{ We have confirmed the existence of  KBHsAH for various $f_a$ ranging from $0.02$ to $10$.}
 of the axion potential on the hairy BHs by performing a thorough study of the solutions with the specific decay constant $f_a = 0.05$.

\section{Numerical Results}

\subsection{The domain of existence}

At the end of the previous section, we fixed two of the four input parameters of the problem ($f_a,m$). Thus, the full domain of existence is obtained by varying the remaining two input parameters: the angular frequency of the scalar field, $\omega$; and the radial coordinate of the event horizon, $r_H$. Since it is impossible to obtain all possible BH solutions, we obtained a very large number of them ($\sim$ 30000) and we have extrapolated this large discrete set of solutions into the continuum, which defines the region where one can find the BH solutions with axionic hair.

\begin{figure}[ht!]
	\begin{center}
		\includegraphics[height=.22\textheight, angle =0]{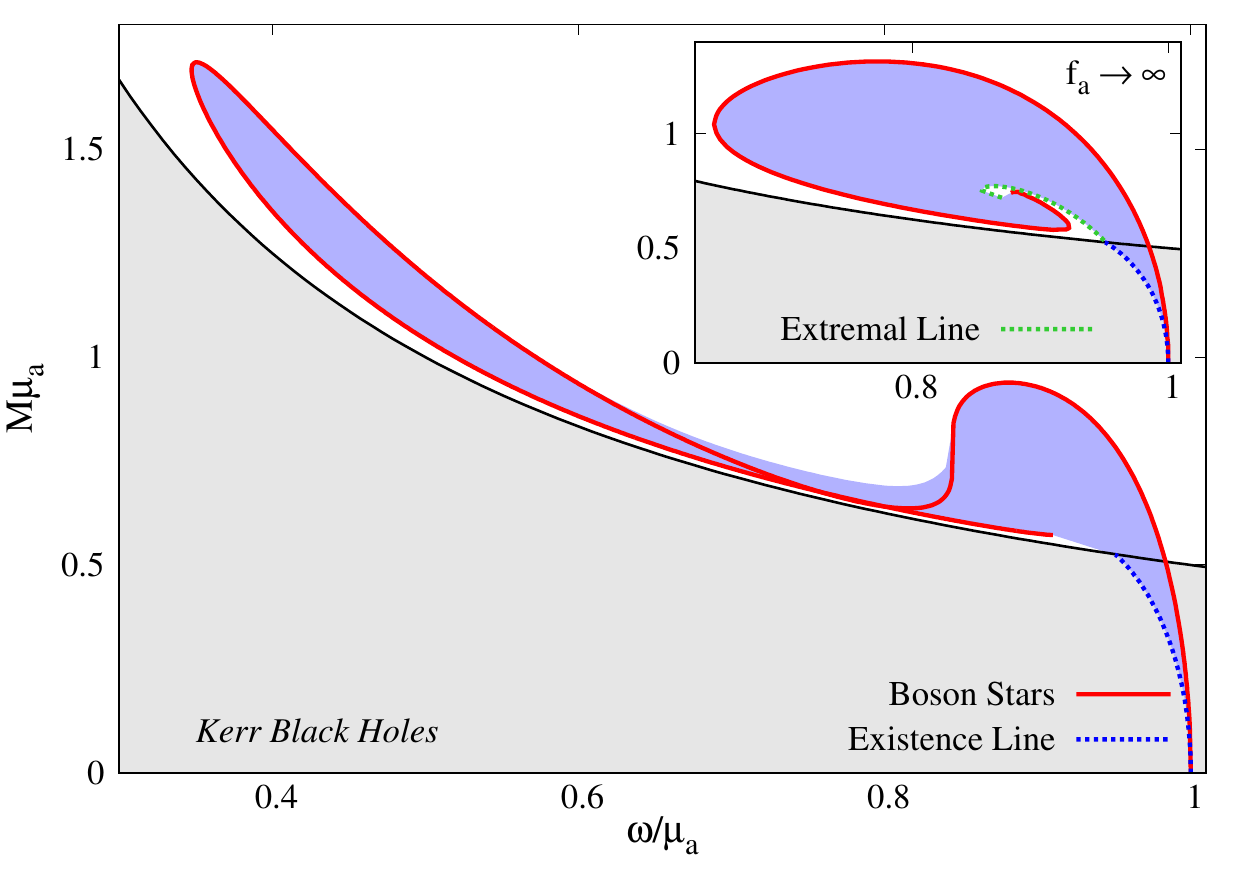}
		\includegraphics[height=.22\textheight, angle =0]{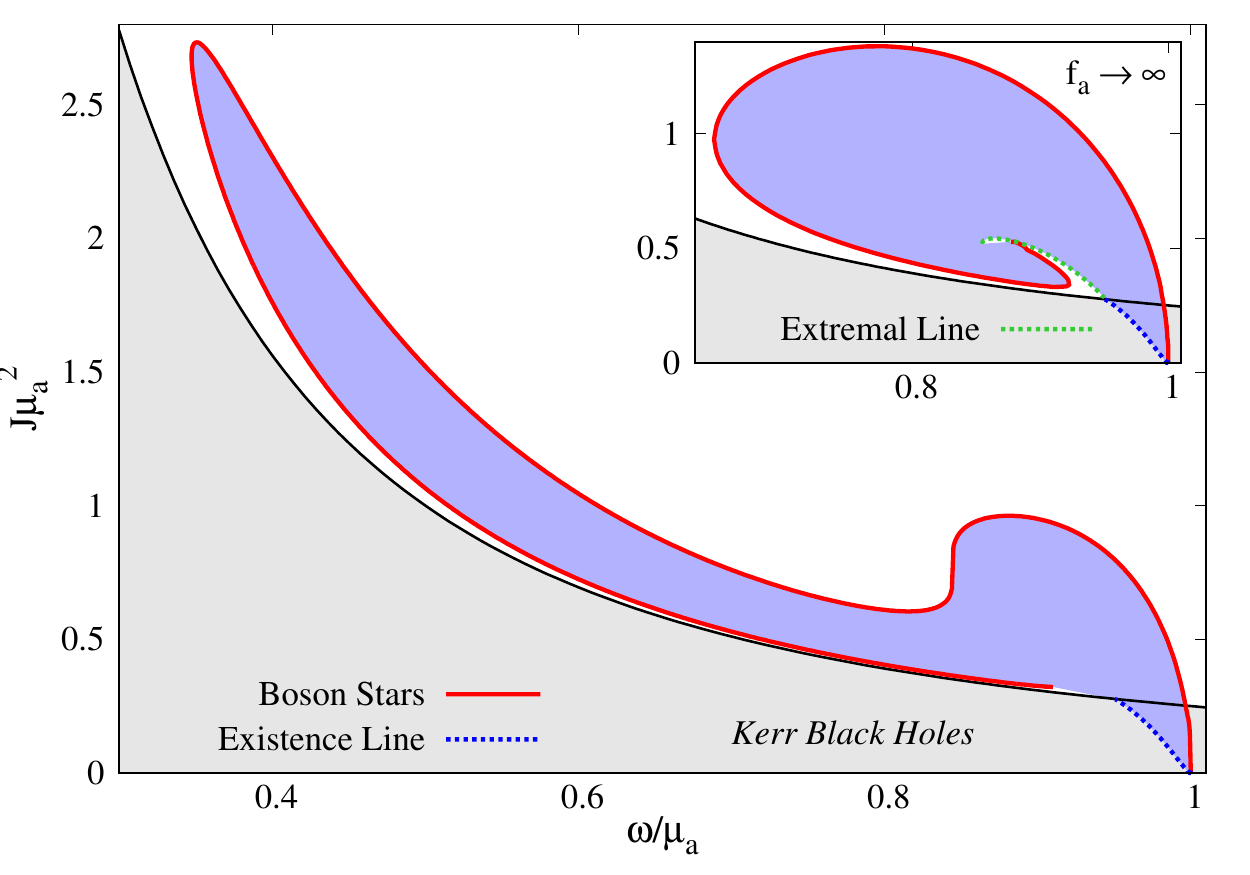}
	\end{center}
	\caption{Domain of existence of KBHsAH for $f_a = 0.05$ in the $M\mu_a$ \textit{vs.} $\omega/\mu_a$ plane (left panel) and in the $J\mu_a^2$ \textit{vs.} $\omega/\mu_a$ plane (right panel). In both panels the insets correspond to the analogous domain of existence for the original Kerr BHs  with synchronised hair (no scalar self-interactions), corresponding  to $f_a \rightarrow \infty$. For the latter family of solutions we also present the extremal line, composed of BHs with vanishing Hawking temperature.}
	\label{Fig:MassAngularMomentum}
\end{figure}

Such a region can be expressed and plotted in various ways. In the left panel of Fig. \ref{Fig:MassAngularMomentum} (main panel), we show it in an ADM mass $M\mu_a$ \textit{vs.} angular frequency $\omega/\mu_a$ plane. We can observe that \textit{most} (but not all) of the numerical solutions region is bounded by two specific lines\footnote{In fact, there is a third line corresponding to extremal hairy BHs, which possess vanishing Hawking temperature. In this work we have only studied the neighbouring solutions of these extremal BHs, but not the latter \textit{per se}, which would require a different metric ansatz.},
\begin{itemize}
	\item The \textit{axion boson star line} - corresponding to the solitonic limit, in which both the event horizon radius and area vanish, $r_H = 0$ and $A_H = 0$, and the solutions have no BH horizon; therefore $q = 1$. Such line is represented in both panels of Fig. \ref{Fig:MassAngularMomentum} as a red solid line.
	
	\item The \textit{existence line} - corresponding to specific subset of vacuum Kerr BHs which can support stationary scalar clouds (with an infinitesimally small $\phi$), first discussed by Hod~ \cite{Hod:2012px,Hod:2013zza}, thus having $q = 0$. These solutions are obtained by linearising the theory, and since, on that regime, the axion self-interacting potential reduces to the mass potential -- \textit{cf.} Eq. \eqref{Eq:AxionPotentialExpansion} -- the existence line will be the same as the one obtained for the family of Kerr BHs with scalar hair. Such line is represented in both panels of Fig. \ref{Fig:MassAngularMomentum} as a blue dotted line.
\end{itemize}

Fig.~\ref{Fig:MassAngularMomentum} (left panel) exhibits a novel property of this class of BHs: the solutions' region is no longer totally bounded by the boson star line in this particular representation. For large decay constant, $f_a$, we recover the family of Kerr BHs with synchronised scalar hair~\cite{Herdeiro:2014goa} (inset), for which the solutions' region \textit{is} totally bounded by the boson star line (together with an existence line -- scalar clouds --, and extremal line -- zero temperature BHs).  The consequence of this observation is that for certain frequencies, the ADM mass  is not maximised by a boson  star, but rather  by a hairy BH.

By changing the representation, however, and plotting the domain of existence in the ADM angular momentum $J\mu_a^2$ \textit{vs.} angular frequency $\omega/\mu_a$ plane -- right panel of Fig. \ref{Fig:MassAngularMomentum} --, we see that, for the angular momentum, the boson star line bounds all axionic BHs with decay constant $f_a = 0.05$. Thus for all frequencies, the angular momentum  is maximised by a boson  star.

Another distinctive feature of the domain of solutions of the axionic BHs is the existence of a local maximum for the mass and angular momentum at $\omega \sim 0.9$, which is not the global maximum. For solutions with smaller angular frequency, it is possible to have BHs with more mass and angular momentum than the ones near the local maximum; in fact, these quantities  are maximised for the solutions with the smallest possible value of angular frequency. Such is not the case in the absence of the axionic potential (inset of both panels in Fig. \ref{Fig:MassAngularMomentum}).

\begin{figure}[ht!]
	\begin{center}
		\includegraphics[height=.22\textheight, angle =0]{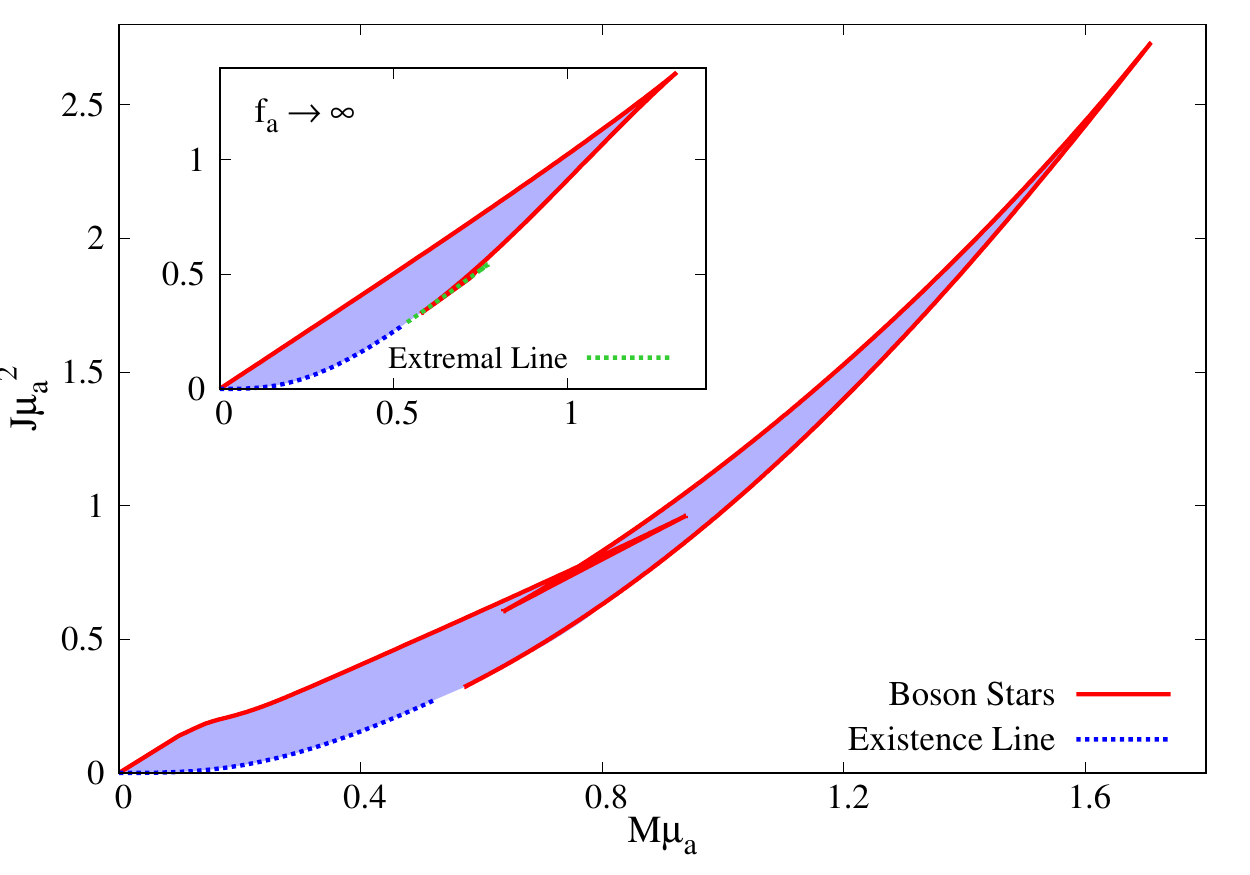}
		\includegraphics[height=.22\textheight, angle =0]{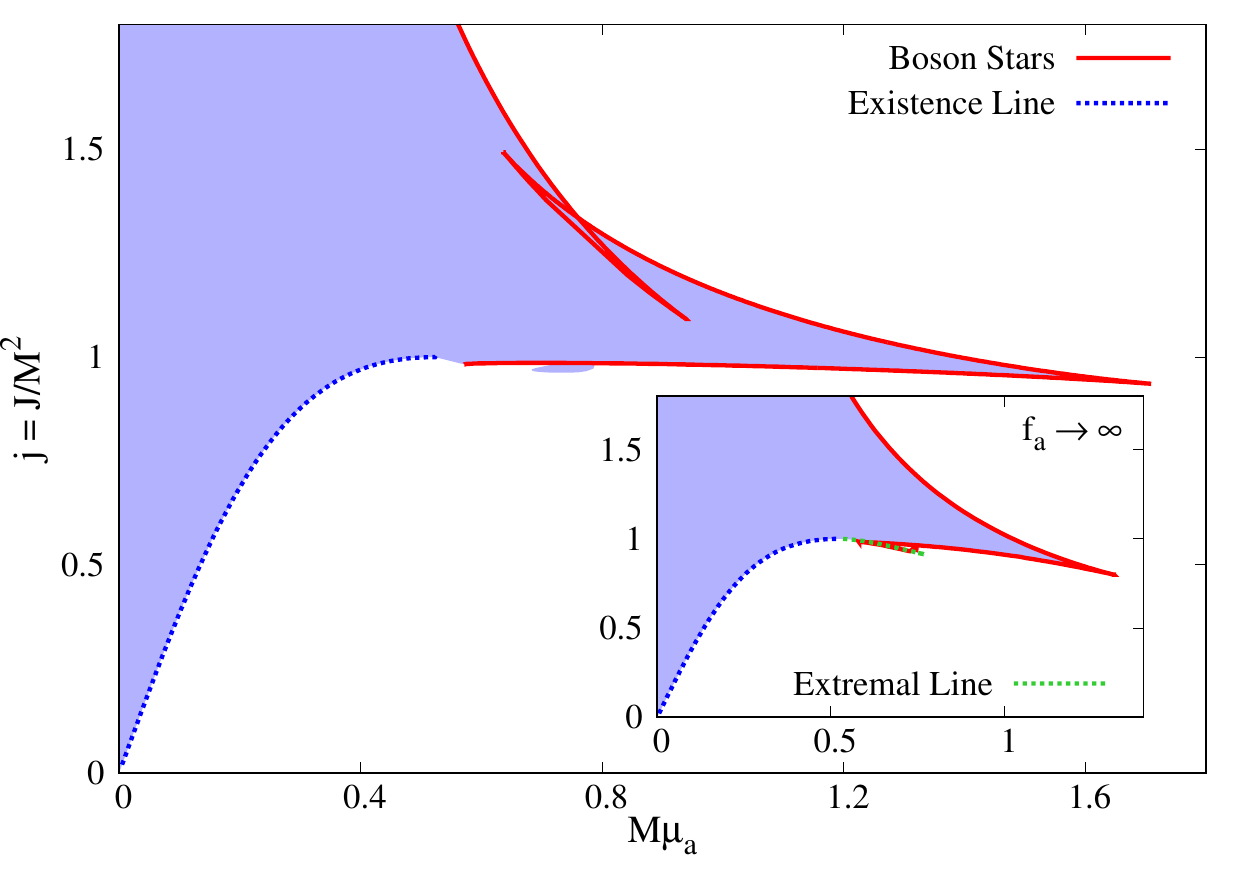}
	\end{center}
	\caption{ADM angular momentum, $J$, (left panel) and dimensionless spin, $j = J/M^2$, (right panel) as a function of the ADM mass $M$. As in Fig.~\ref{Fig:MassAngularMomentum} the insets show the free scalar field limit $f_a \rightarrow \infty$, where the extremal line is also shown.}
	\label{Fig:PhaseSpaceSpin}
\end{figure}

In Fig. \ref{Fig:PhaseSpaceSpin}  both the ADM angular momentum, $J$ and the dimensionless spin, $j$, defined as $j \equiv J/M^2$, are exhibited  $vs.$ the ADM mass, $M$, on the left and right panels, respectively.
In the former, we can see that the axionic BHs can have a higher mass and angular momentum than their $f_a \rightarrow \infty$ counterparts (inset).  We can also see a zigzag of the boson star line. This behaviour is explained by the sudden drop on the ADM mass and angular momentum around $\omega/\mu_a \approx 0.84$ -- \textit{cf.} Fig. \ref{Fig:MassAngularMomentum}.
In the right  panel, we see a considerable violation of the Kerr bound, $j \leqslant 1$ in part  of solution space. This already occurred for the $f_a \rightarrow \infty$ limit (inset). Again, it is possible to visualise the zigzag behaviour of the boson star line.

\begin{figure}[ht!]
	\begin{center}
		\includegraphics[height=.22\textheight, angle =0]{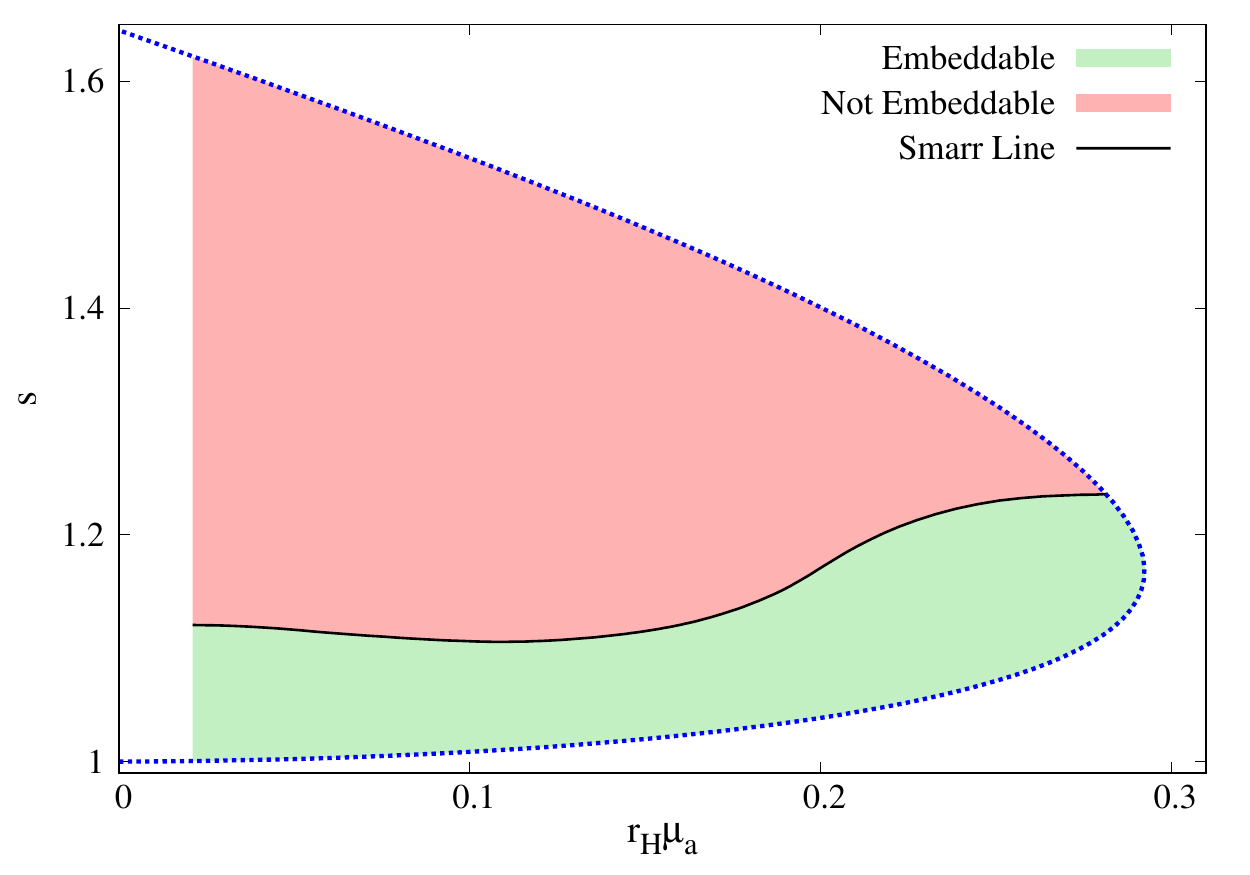}
		\includegraphics[height=.22\textheight, angle =0]{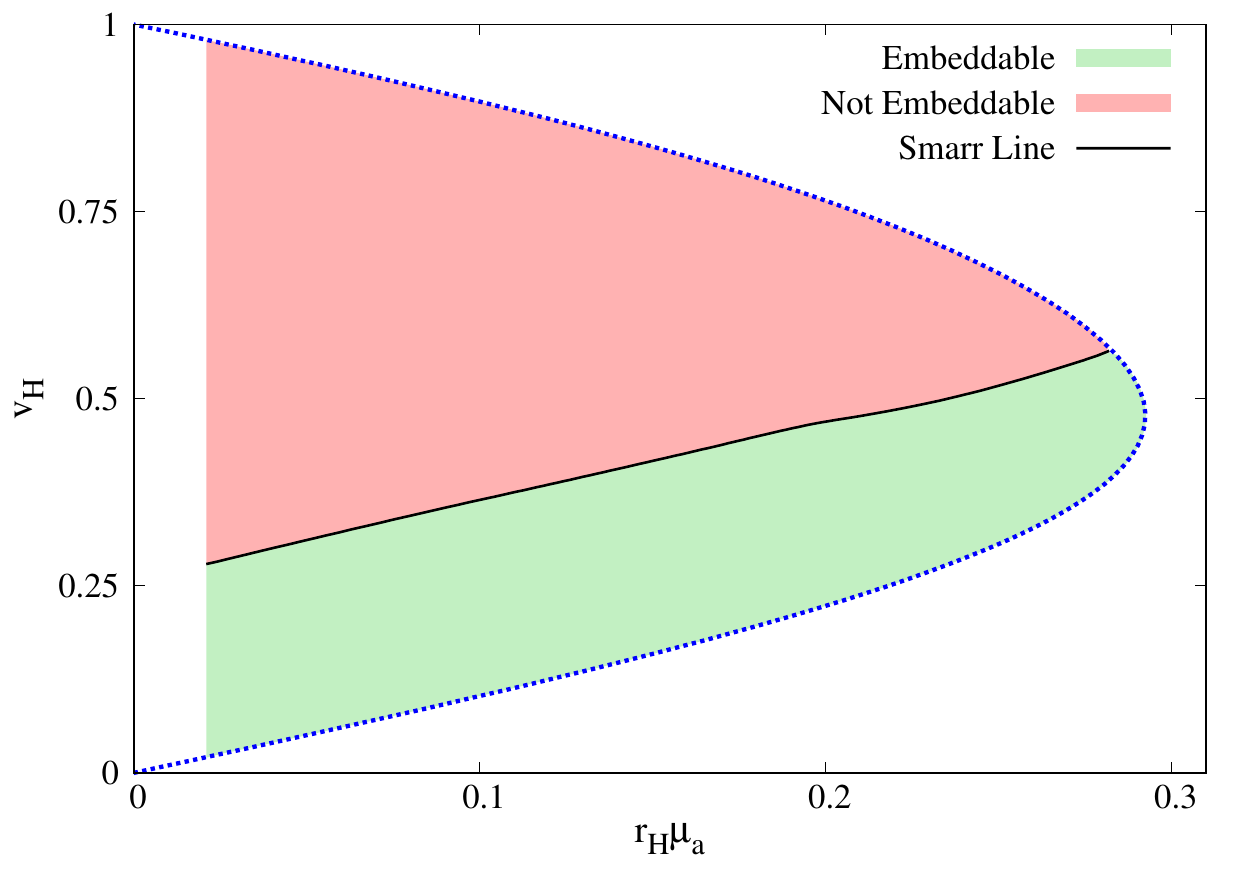}
	\end{center}
	\caption{The sphericity, $\mathfrak{s}$ (left panel), and the horizon linear velocity, $v_H$ (right panel), as a function of the event horizon radial coordinate, $r_H \mu_a$. The black solid line corresponds to the Smarr line, which, in contrast with the $f_a \rightarrow \infty$ limit, is no longer constant for the sphericity. Below (above) it, we have (do not have) embeddable BH horizons in Euclidean 3-space. The white region in the far left region of each panel corresponds to solutions outside  our scan (challenging due to the small $r_H$). }
	\label{Fig:SphevH}
\end{figure}

Let us  now study the horizon geometry of the axionic BHs. Their event horizon  has a spherical topology but a spheroidal geometry, similarly to Kerr BHs. This can be seen by studying the spatial cross-section of the horizon, through the induced metric,
\begin{equation}
	d\Sigma^2 = r_H^2 \left[ e^{2 F_1(r_H,\theta)} d\theta^2 + e^{2F_2(r_H,\theta)} \sin^2 \theta d \varphi^2 \right] \hspace{2pt} .
\end{equation}
Due to the rotation of the solutions, the horizon is squashed at the poles. To show this, we compute the horizon circumference along the equator, $L_e$, and along the poles, $L_p$,
\begin{equation}
	L_e = 2\pi r_H e^{F_2 (r_H, \pi/2)} \hspace{2pt}, \hspace{10pt} L_p = 2 r_H \int_0^\pi d\theta e^{F_1(r_H,\theta)} \hspace{2pt}.
\end{equation}
We define the sphericity as the ratio of both circumferences above \cite{Delgado:2018khf},
\begin{equation}
	\mathfrak{s} = \frac{L_e}{L_p}
\end{equation}
For values in which the sphericity is greater (lower) than 1, the horizon will be squashed (elongated) at the poles, leading to an oblate (prolate) spheroid. From the left panel in Fig. \ref{Fig:SphevH} we see that all solutions have a sphericity larger than the unity; thus all solutions have an oblate horizon, as expected. 

Another physical quantity of interest associated with the horizon is its linear velocity $v_H$ \cite{Delgado:2018khf,Delgado:2019prc,Herdeiro:2015moa}. Such quantity measures how fast the null geodesics generators of the horizon rotate relatively to a static observer at spatial infinity. Its definition is quite simple, only taking into account the perimetral radius of the circumference located at the equator, $R_e \equiv L_e/2\pi$, and the horizon angular velocity, $\Omega_H$,
\begin{equation}
	v_H = \frac{L_e}{2\pi} \Omega_H \hspace{2pt} .
\end{equation}
The horizon linear velocity is presented in the right panel of Fig. \ref{Fig:SphevH}. The central feature in this plot is the fact that all solutions have horizon linear velocity smaller than the unity, which, in the units we are using, corresponds to the speed of light. Therefore, null geodesics generators of the horizon never rotate relatively to the asymptotic observer at superluminal speeds, even though some solutions strongly violate the Kerr bound $j \leqslant 1$.

A final insight about the horizon geometry of the axionic BHs is obtained from investigating whether an isometric embedding of the spatial sections of the horizon is possible in Euclidean 3-space $\mathbb{E}^3$. For a Kerr BH, such embedding is possible iff its dimensionless spin obeys $j \leqslant j^{(S)}$~\cite{Smarr:1973zz}, where $ j^{(S)} \equiv  \sqrt{3}/2$ was dubbed the \textit{Smarr point}~\cite{Delgado:2018khf}. For $j > j^{(S)}$ the Gaussian curvature of the horizon becomes negative in the vicinity of the poles \cite{Smarr:1973zz}, which, prevents the embedding (due to occurring at a fixed  point  of the axi-symmetry). 
As expected, this feature also occurs for the axionic BHs. Due to the existence of scalar hair around the BH, we have an extra degree of freedom, which converts the Smarr point into a \textit{Smarr line}. Such line is represented in both panels of Fig. \ref{Fig:SphevH} as a solid black line. One observes that, for both the sphericity and the horizon linear velocity, the Smarr line is not constant. This  contrasts with the behaviour for $f_a \rightarrow \infty$; in that case, the sphericity of the Smarr line was constant and equal to the value of the Smarr point in Kerr \cite{Delgado:2018khf}. Thus, the axion potential destroys the constancy of the sphericity along the Smarr line.

\subsection{Other properties}

\subsubsection{Ergo-regions}

An ergo-region is a part of a spacetime, outside the event horizon, wherein the norm of the asymptotically timelike Killing vector $\xi = \partial_t$ becomes positive and thus the vector becomes spacelike. Ergoregions are associated with the possibility of energy extraction from a spinning BH, via the Penrose process \cite{Penrose:1969pc,Penrose:1971uk}, or superradiant scattering~\cite{Brito:2015oca}. In the context of BHs with synchronised hair, the superradiant instability of vacuum   Kerr BHs in the presence of ultralight  scalar fields is one of the possible channels of formation  of these hairy BHs~\cite{East:2017ovw,Herdeiro:2017phl}. Thus, it is of relevance to analyse ergo-regions for the axionic BHs.

\begin{figure}[ht!]
	\begin{center}
		\includegraphics[height=.28\textheight, angle =0]{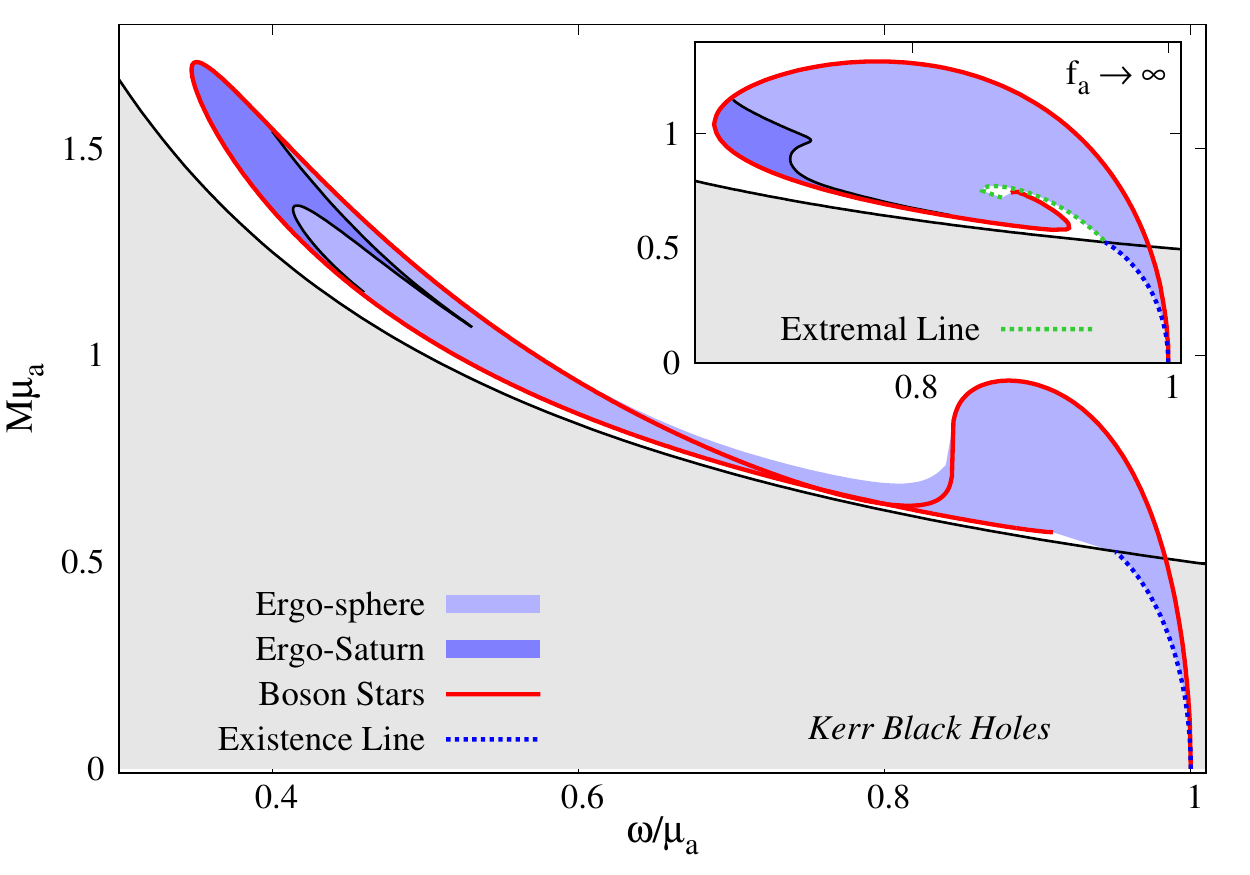}
	\end{center}
	\caption{Ergo-regions. BHs with axionic hair have an ergo-sphere in the light blue region and an ergo-Saturn in the dark blue region. The inset shows the free scalar field case ($f_a \rightarrow \infty$), for comparison.}
	\label{Fig:Ergoregions}
\end{figure}

Kerr BHs possess an ergo-region whose boundary has a spherical topology and touches the BH horizon at the poles -- such surface is called an \textit{ergo-sphere}. For the axionic BHs in the $f_a \rightarrow \infty$ limit, the ergo-regions can be more complicated~\cite{Herdeiro:2014jaa}. Some solutions have a Kerr-like ergo-region; but others have a more elaborate ergo-region topology, with two disjoint parts, one  Kerr-like and another of toroidal topology. The latter  were dubbed \textit{ergo-Saturns}. The toroidal ergo-region is inherited from the mini boson star environment around the horizon, since these stars develop such ergo-regions,  when  sufficiently compact. Such ergo-torii also occur for spinning axion boson stars~\cite{Delgado:2020udb}. Thus, we expect some axionic BHs to develop an ergo-Saturn. This is confirmed in Fig. \ref{Fig:Ergoregions}.
 We have found that the ergo-region  of axionic BHs follows a qualitatively similar distribution to that of their $f_a \rightarrow \infty$ limit: there are solutions which possess an ergo-sphere and others that develop an ergo-Saturn.  The latter occur  on the far left of the domain of existence in Fig. \ref{Fig:Ergoregions}, where the most massive BHs exist, having the lowest possible values of the angular frequency of the scalar field.

\subsubsection{Light rings and timelike innermost stable circular orbits}

A phenomenological aspect of importance is the structure of circular orbits (COs) of both massless and massive particles around a BH. In particular, the (timelike) innermost stable circular orbit (ISCO) and the (null) light rings (LRs) are of special relevance. The former is associated with a cut-off frequency of the emitted synchrotron radiation generated from accelerated charges in accretion disks; the latter is related to the real part of the frequency of BH quasi-normal modes \cite{Cardoso:2008bp}, as well as to the BH shadow \cite{Cunha:2018acu}.

The structure of COs can be obtained as follows. Given the geometry in Eq. \eqref{Eq:AnsatzMetric}, we can compute the effective Lagrangian for equatorial, $\theta = \pi/2$, geodesic motion as,
\begin{equation}\label{Eq:EffectiveLagrangian}
	2\mathcal{L} = \frac{e^{2 F_1}}{N}\dot{r}^2 + e^{2 F_2}r^2 \left( \dot{\varphi} - W \dot{t} \right)^2 - e^{2F_0} N \dot{t} = \epsilon \ ,
\end{equation} 
where all \textit{ansatz} functions depend only on the radial coordinate $r$, the dot, $\dot{ }$, denotes the derivative \textit{w.r.t} the proper time, and $\epsilon = \{-1,0\}$ for massive (timelike) particles and massless (lightlike) particles, respectively. Due to the existence of two Killing vector fields, we can write $\dot{t}$ and $\dot{\varphi}$ in terms of the energy $E$ and angular momentum $L$ of the particle,
\begin{eqnarray}
	&&E = \left( e^{2F_0} N - e^{2F_2} r^2 W^2 \right) \dot{t} + e^{2F_2} r^2 W \dot{\varphi} \ , \\
	&&L = e^{2 F_2} r^2 \left( \dot{\varphi} - W \dot{t} \right) \ .
\end{eqnarray}
Inverting the above system of equations and replacing the result into the effective Lagrangian, Eq. \eqref{Eq:EffectiveLagrangian}, we can obtain an equation for $\dot{r}$, which defines a potential $V(r)$,
\begin{equation}
	\dot{r}^2 = V(r) \equiv e^{-2 F_1} N \left[ \epsilon + \frac{e^{2F_0}}{N} \left( E - L W \right)^2 - e^{2F_2} \frac{L^2}{r^2} \right] \ .
\end{equation}

In order to obtain COs, both the potential and its derivative must be zero, \textit{i.e.} $V(r) = V'(r) = 0$. Depending on whether we are considering massless or massive particles, these two equations will yield different results. 

For massless particles, the first equation, $V(r) = 0$, will give two algebraic equation for the impact parameter of the particle, $b_+ \equiv L_+/E_+$ and $b_- \equiv L_-/E_-$, corresponding to co- and counter-rotating orbits, respectively. The second equation, $V'(r) = 0$, together with the impact parameters, will give the radial coordinate of the co- and counter-rotating LRs. Whenever it is possible to obtain a real solution for the radial coordinate, the BH possesses LRs. 

\begin{figure}[ht!]
	\begin{center}
		\includegraphics[height=.22\textheight, angle =0]{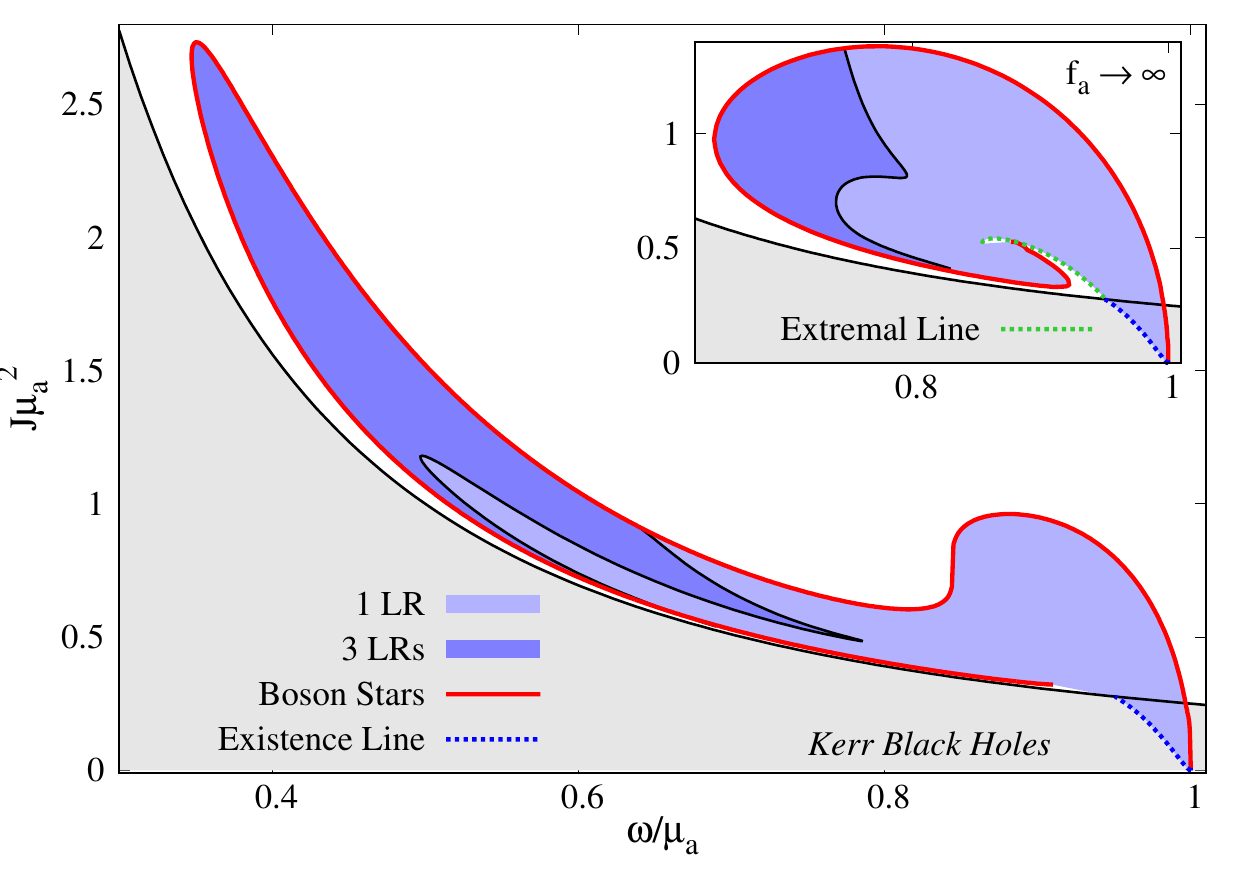}
		\includegraphics[height=.22\textheight, angle =0]{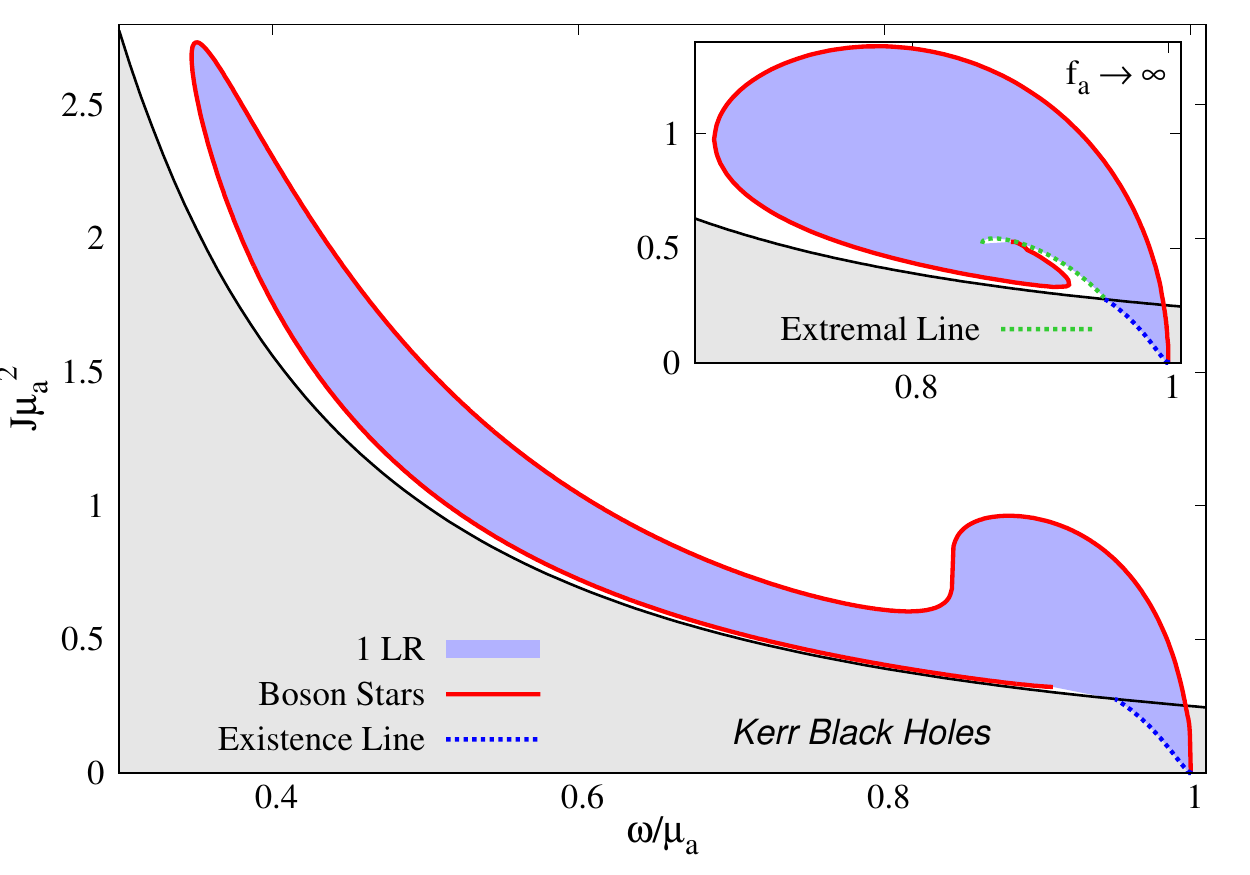}		
	\end{center}
	\caption{Number of LRs. The left (right) panel shows the counter-rotating (co-rotating) case. Hairy BHs can have 1 LR, as for the case of Kerr, in the light blue region, or 3 LRs, in the dark blue region. The inset follows the same description but for $f_a \rightarrow \infty$.}
	\label{Fig:LRs}
\end{figure}

\begin{figure}[ht!]
	\begin{center}
		\includegraphics[height=.22\textheight, angle =0]{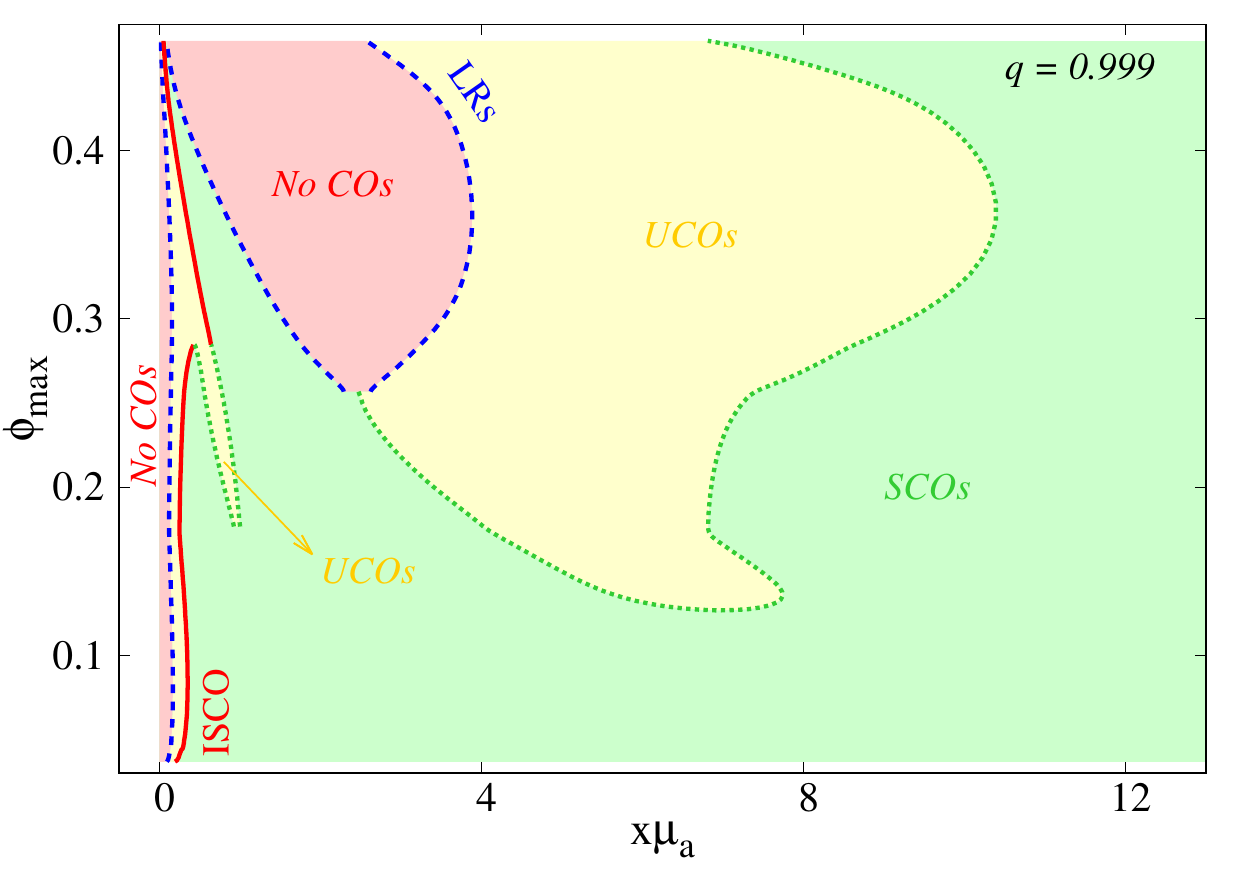}
		\includegraphics[height=.22\textheight, angle =0]{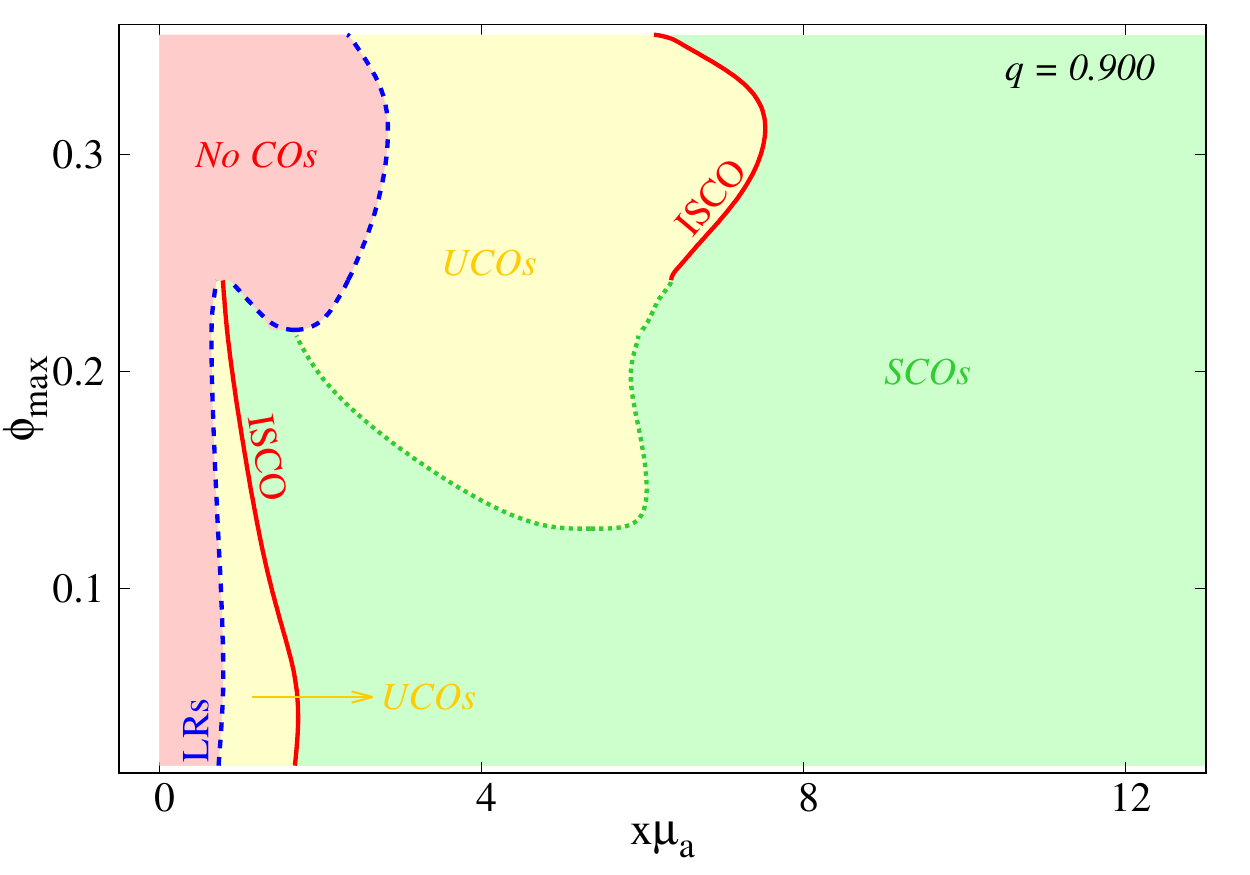} \\
		\includegraphics[height=.22\textheight, angle =0]{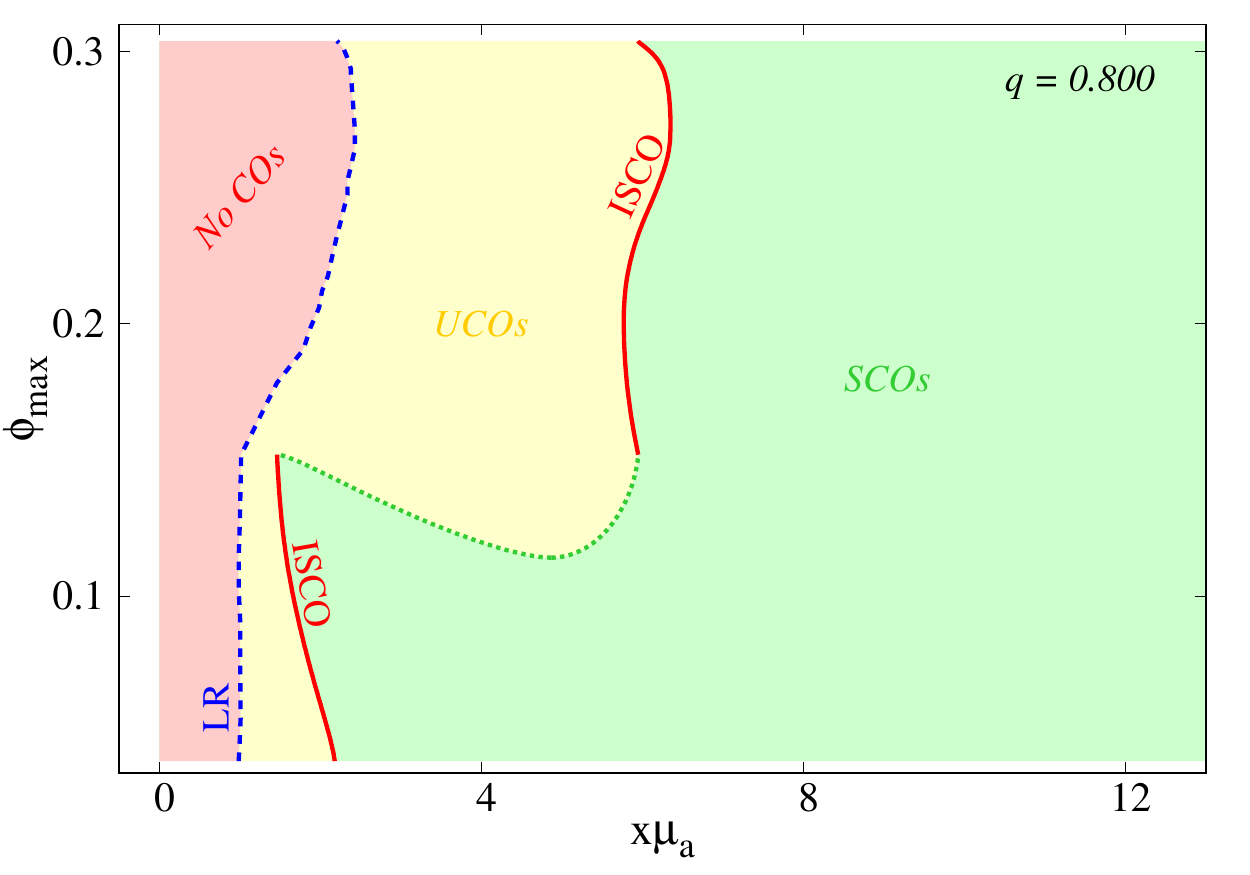}
		\includegraphics[height=.22\textheight, angle =0]{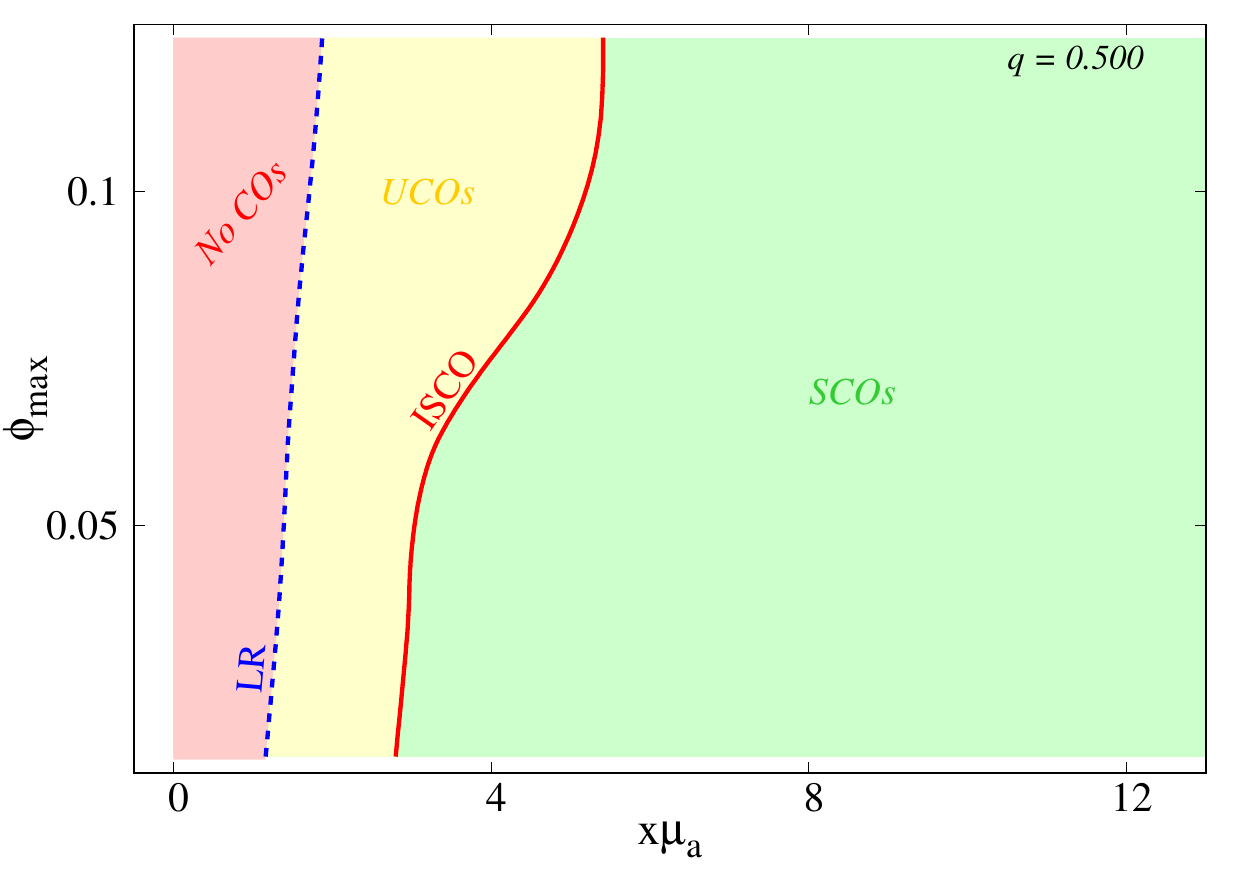}
	\end{center}
	\caption{Structure of counter-rotating COs for four sets of axionic BHs with constant $q = \{0.5, 0.8, 0.9, 0.999\}$. The maximal value of the scalar field $\phi_\text{max}$ varies with $q$; thus the vertical scale changes for the four plots.}
	\label{Fig:CounterCOs}
\end{figure}

\begin{figure}[ht!]
	\begin{center}
		\includegraphics[height=.26\textheight, angle =0]{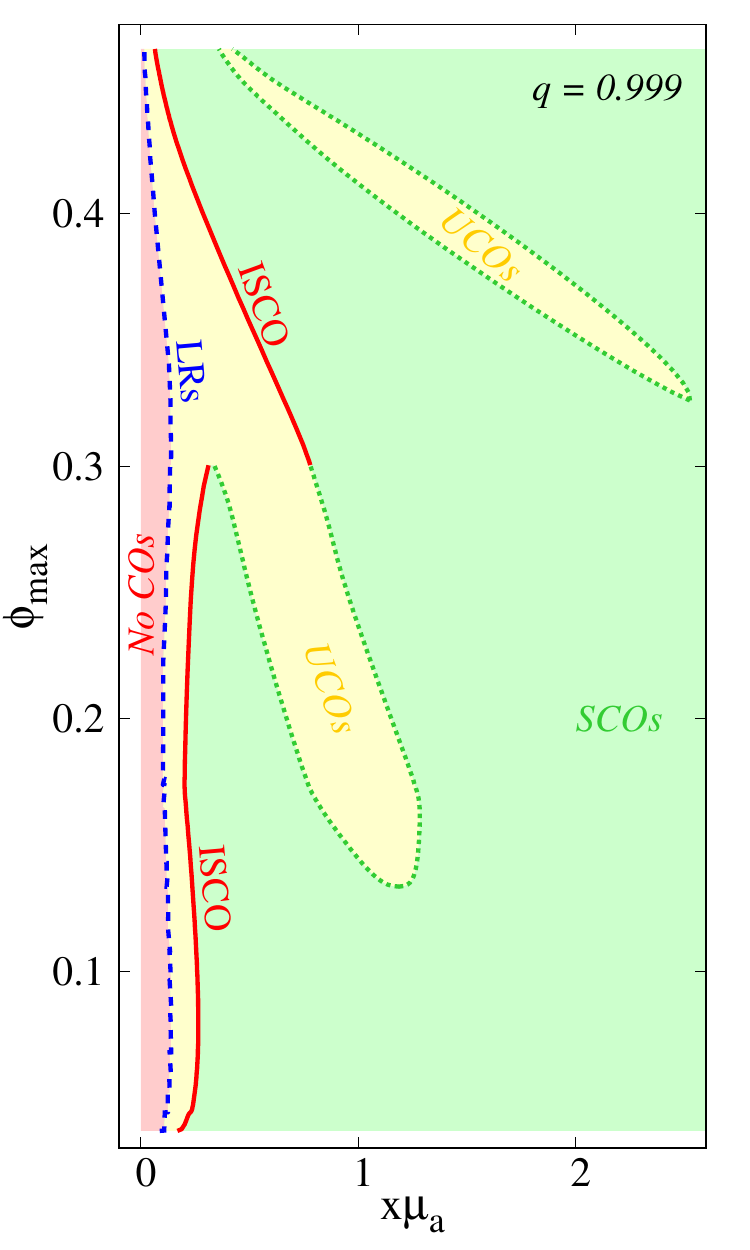}
		\includegraphics[height=.26\textheight, angle =0]{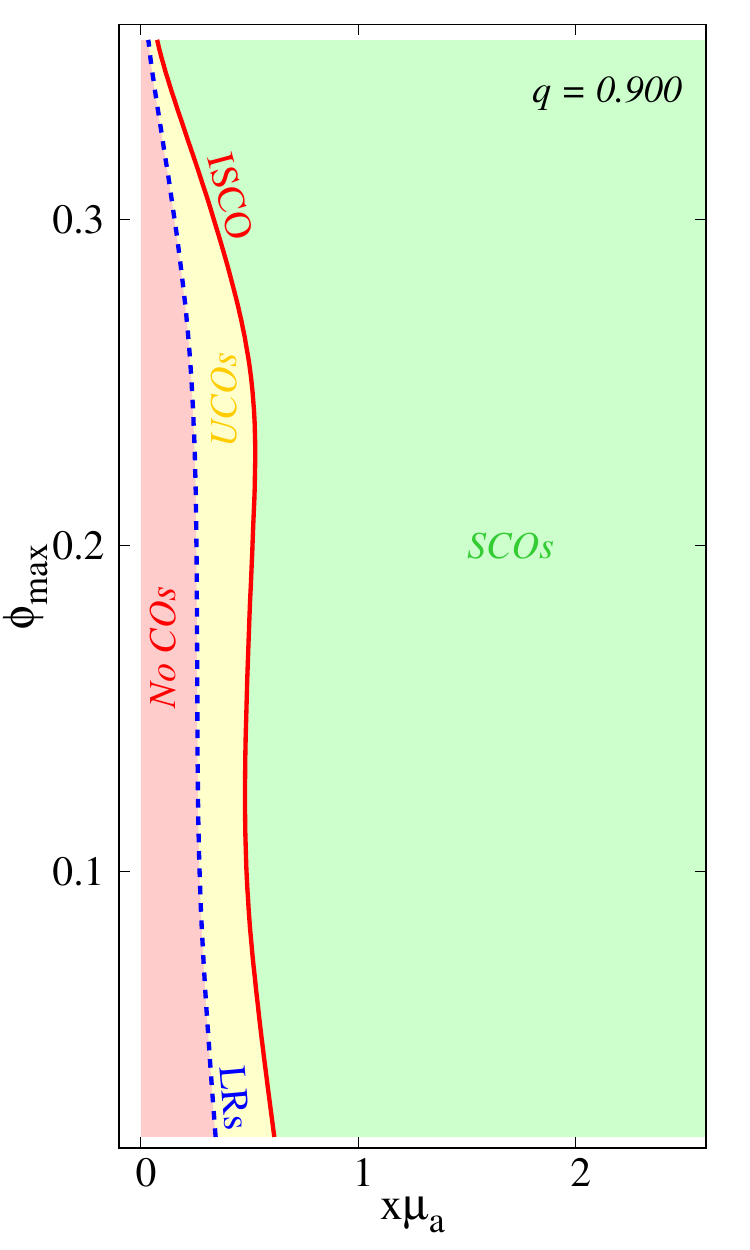} 
		\includegraphics[height=.26\textheight, angle =0]{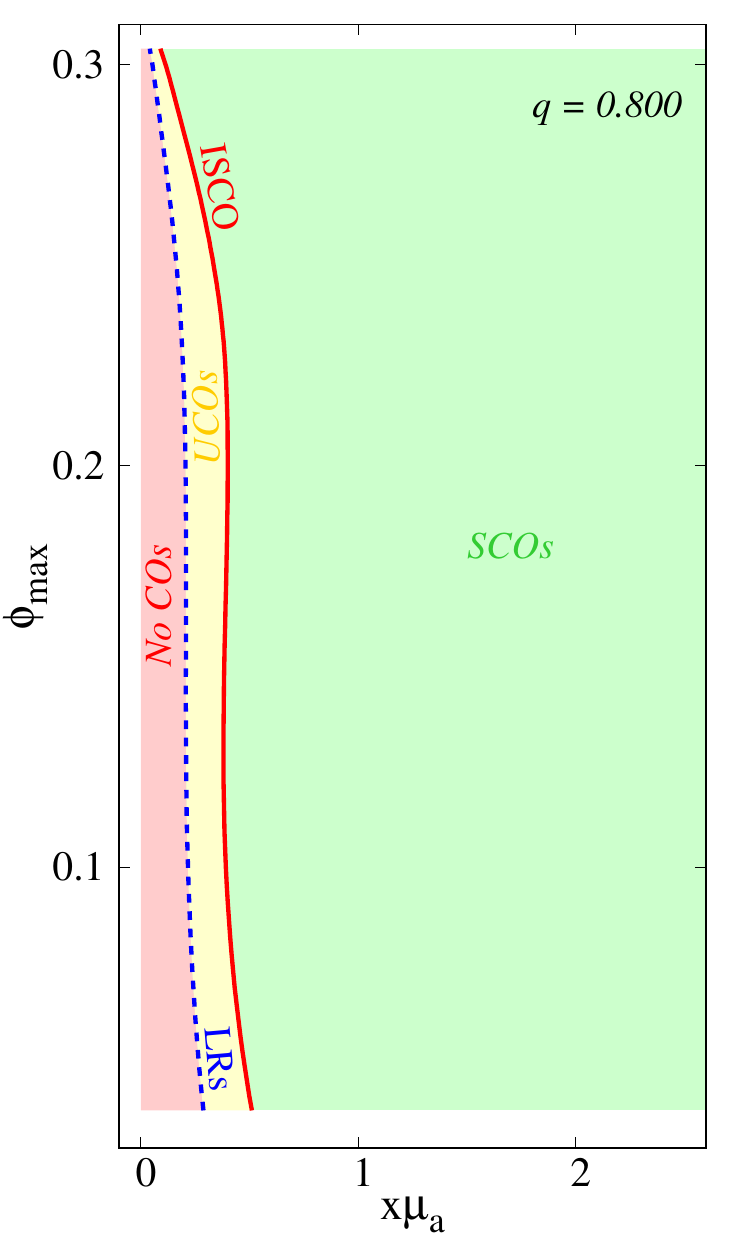}
		\includegraphics[height=.26\textheight, angle =0]{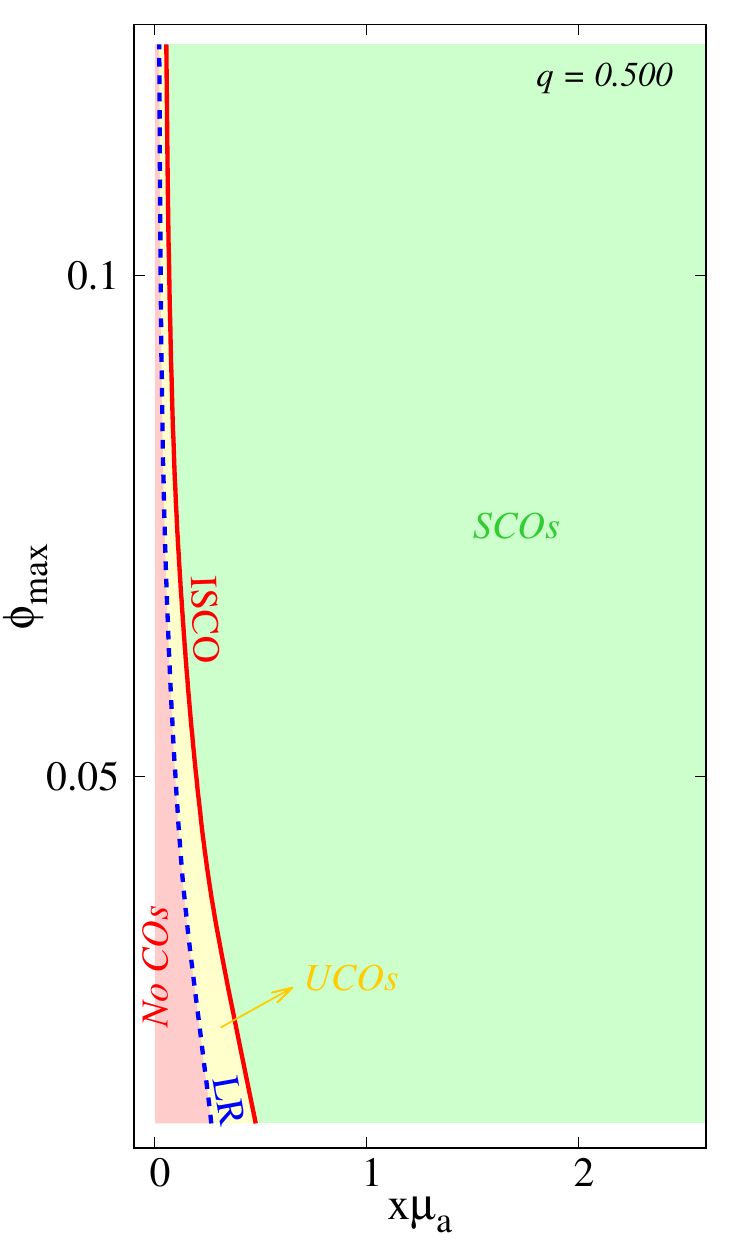}
	\end{center}
	\caption{Structure of co-rotating COs for four sets of axionic BHs with constant $q = \{0.5, 0.8, 0.9, 0.999\}$. Note that the maximal value of the scalar field $\phi_\text{max}$ varies with $q$, thus the vertical scale changes for the four plots.}
	\label{Fig:CoCOs}
\end{figure}

In Fig. \ref{Fig:LRs}, we show the distribution of hairy BHs with different number of LRs in the angular frequency, $\omega/\mu_a$ \textit{vs.} angular momentum, $J \mu^2$ plane.  The left panel shows the counter-rotating case, whereas the right panel shows the co-rotating case. 
In both cases, it is always possible to have at least one LR, as for the Kerr BH. In the counter-rotating case,   however, if the surrounding scalar field is compact enough, an extra pair of LRs emerge, leading to a hairy BH with 3 LRs. This is the case of a large set of hairy BHs with low values of $\omega/\mu_a$. In fact, we see the same behaviour for the free  scalar field case (inset plot); even the solid black line separating the two regions has a qualitatively similar shape. 
The radial coordinate of the several LRs is  shown in Figs. \ref{Fig:CounterCOs} and \ref{Fig:CoCOs} as a blue dashed line.

For massive particles, $V(r) = V'(r) = 0$ will yield two algebraic equations for the energy and angular momentum of the particle, $\{E_+,L_+\}$ and $\{E_-,L_-\}$ corresponding to co- and counter-rotating orbits, respectively. Their stability can be verified by analysing the sign of the second derivative of the potential $V(r)$. Given a BH solution it will have three distinct regions concerning (timelike) COs:
\begin{itemize}
	\item \textit{No circular orbits (No COs)} - Whenever we obtain an imaginary solution for the energy and angular momentum of the massive particle.  This is a region in which no COs exist. Such region is shaded in light red, in Figs. \ref{Fig:CounterCOs} and \ref{Fig:CoCOs}.
 	
	\item \textit{Unstable circular orbits (UCOs)} - In this region, it is possible to obtain real solutions for the energy and angular momentum, but the sign of the second derivative of the potential $V(r)$ is positive, implying that COs are unstable. Such region is shaded in light yellow, in Figs. \ref{Fig:CounterCOs} and \ref{Fig:CoCOs}.
	
	\item \textit{Stable circular orbits (SCOs)} - In this region, it is also possible to obtain real solutions for the energy and angular momentum, but now the second derivative of the potential $V(r)$ has a negative sign, thus implying that COs are stable. Such region is shaded in light green, in Figs. \ref{Fig:CounterCOs} and \ref{Fig:CoCOs}.
\end{itemize}

The ISCO, as the name entails, is the innermost stable circular orbit, \textit{i.e.}, the  stable CO with the smallest radial coordinate. At this orbit, the second derivative of the potential vanishes, as it corresponds to the transition between the SCOs region and the UCOs region. The ISCO is represented as a solid red line in Figs. \ref{Fig:CounterCOs} and \ref{Fig:CoCOs}.

In Fig.~\ref{Fig:CounterCOs} we present the structure of counter-rotating COs for four sets of hairy BHs with constant $q = \{0.5, 0.8, 0.9, 0.999\}$. For each set of BHs, we have constructed a plot showing the maximal value of the scalar field, $\phi_\text{max}$, as a function of the radial coordinate $x = \sqrt{r^2 - r_H^2}$. This ensures that every horizontal line (lines with constant $\phi_\text{max}$) corresponds to a unique BH solution making it easier to analyse the structure of COs.

The first observation from  Fig.~\ref{Fig:CounterCOs} is that, by increasing $q$, it is possible to obtain solutions with larger $\phi_\text{max}$. This is reasonable since by increasing $q$ the BHs become hairier. Let us now analyse each individual plot. For the case where $q = 0.5$ (right bottom panel), the structure of the counter-rotating COs for all BHs is identical to the one for Kerr. There is only one LR and one ISCO, which separates the No COs region from the UCOs region, and separates the UCOs region from the SCOs region, respectively. At large enough $x$ we only have SCOs, but as we approach the horizon, we reach the ISCO and consequently enter the UCOs region. If we continue towards the horizon, we eventually cross the LR and enter the No COs region.
The radial coordinate of the LR and ISCO increases monotonically with the increase of the maximal value of the scalar field.

For the case where $q = 0.8$ (left bottom panel) most of the structure is similar, but a new feature emerges. Starting at the regime where there are BHs with a dilute scalar field ($\phi_\text{max} < 0.11$), their structure is the same as for Kerr, but for BHs with $\phi_\text{max} > 0.11$ and $\phi_\text{max} < 0.15$  an extra region of UCOs appears besides the one already present between the LR and the ISCO, and  disconnected from the latter. The new region appears as a single point around $x\mu_a \approx 4.8$ and increases in size as $\phi_\text{max}$ increases until its inner boundary merges with the ISCO. For $\phi_\text{max} > 0.15$, the structure is again the same as that of Kerr, but now the radial coordinate of the ISCO is significantly larger than for BHs with a more dilute scalar field. In fact, there is a discontinuity when we study the evolution of the radial coordinate of the ISCO as it changes with the maximum value of the scalar field, as one can see in the left bottom panel of Fig. \ref{Fig:CounterCOs}.

In the $q = 0.9$ case (right top panel), we have a structure similar as for the $q = 0.8$ when we consider BHs with $\phi_\text{max} < 0.21$, but, for the remaining BHs, the scalar field environment is compact enough to develop a pair of extra LRs. Such pairs give rise to a new region of No COs, disconnected from the already existing one between the event horizon and innermost LR. It starts as a single point for a BH with $\phi_\text{max} \approx 0.21$; then it grows in size until its inner boundary (one of the LRs) merges with the innermost LR, connecting both regions of No COs. At the same time, the ISCO also merges with the LRs, meaning that such BH has a degenerate point in which two LRs and the ISCO converge. For BHs with a larger value of $\phi_\text{max}$, a Kerr-like structure again emerges, but now both the LR and the ISCO appear at a larger radial coordinate than for small $\phi_\text{max}$.

Lastly, in the $q = 0.999$ case (left top panel), the most complex structure is observed. This case inherits the several features discussed in the previous cases, such as the existence of two different and disconnected UCOs and No COs regions, but also presents a new one. For BHs with $\phi_\text{max} > 0.18$, a third region of UCOs between the two already existing ones. This third region starts as a single point and increases slowly in size as $\phi_\text{max}$ increases until its inner boundary connects with the ISCO, and the innermost region of UCOs merges with this new region. Although it is not visible in the left top panel of Fig. \ref{Fig:CounterCOs}, one can draw the conclusion that there is a BH with larger $\phi_\text{max}$ for which the innermost two LRs together with the ISCO converge to the same single point, akin to what happened in the $q=0.9$ case.

Now we turn to the co-rotating COs. In Fig.~\ref{Fig:CoCOs} we followed the same idea that we used to study  the counter-rotating COs and plotted four sets of BHs solutions with constant $q = \{0.5, 0.8, 0.9, 0.999\}$ in the  $\phi_\text{max}$ \textit{vs.} $x$ plane.  
Fig. \ref{Fig:CoCOs} manifests that the structure of co-rotating COs is much simpler than the counter-rotating one; only the  $q=0.999$ case presents significant differences from the other ones;  the structure of the latter is the same as for a Kerr BH. Moreover, the radial coordinate of the co-rotating LR and ISCO is always smaller than the one of the counter-rotating LR and ISCO, as it is for the  Kerr  BH~\cite{Bardeen:1972fi}.

Let us comment on the  qualitatively different $q=0.999$ case (first panel of Fig. \ref{Fig:CoCOs}). For BHs with a dilute scalar field, a Kerr-like structure is  observed; but above $\phi_\text{max} \approx 0.14$  a second region of UCOs  emerges. This region exists until we reach a BH with $\phi_\text{max} \approx 0.30$, where both regions of UCOs merge together. For BHs with slighter larger $\phi_\text{max}$ a second region of UCOs again emerges, but now this region occurs at large radii, quickly decreasing to smaller radii as $\phi_\text{max}$ increases. Although not shown in this panel, we can predict that this new region of UCOs will converge, as the previous one, with the already existing and closer to the event horizon UCOs region.



\section{Conclusions and Remarks}

In this work, we have constructed and analysed BHs with synchronised  axionic hair, which are BH generalisation of the rotating axion boson stars recently found in \cite{Delgado:2020udb} - see also~\cite{Guerra:2019srj}. These are stationary, axi-symmetric, asymptotically flat and regular on and outside the event horizon solutions of the Einstein-Klein-Gordon equations of motion with a QCD axion-like potential, Eq.~\eqref{Eq:AxionPotential}. This family of axionic BHs is described by three parameters: the radial coordinate of the event horizon, $r_H$, the angular frequency of the scalar field, $\omega$ and the decay constant of the QCD potential, $f_a$. In this work, we have thoroughly scanned the space of solutions with decay constant $f_a = 0.05$, since, from the results found in \cite{Delgado:2020udb}, this  yields a case with considerable impact of the axion potential, and hence considerable differences from the free  scalar field case, obtained as the $f_a \rightarrow \infty$  limit.  The latter yields the  original example of  Kerr BHs with synchronised scalar hair \cite{Herdeiro:2014goa}. Even larger deviations from  this original example may occur for even smaller $f_a$, but then the numerics to obtain  such solutions becomes more challenging.

When comparing the axionic BHs with their free scalar field counterparts~\cite{Herdeiro:2014goa},  there are both differences and similarities. Some key differences are: (i) axionic BHs can have more mass and angular momentum and lower values of the angular frequency of the scalar field; (ii) the existence of a local maximum for the mass and angular momentum, that does not exists for the $f_a \rightarrow \infty$ case; (iii) the presence of a small region of frequencies where the mass of axionic BHs is no longer bounded by the axion boson stars. In such region, we have a degeneracy of solutions with the same angular frequency and ADM mass, but such degeneracy is easily lifted by specifying any other physical quantity; (iv)  the variation of the sphericity along the Smarr line, which, in the free scalar field case was constant and equal to the sphericity of the Smarr point for Kerr BHs, but varies for axionic BHs. 

Concerning the similarities, we may emphasise: (i)  the clear violation of the Kerr bound, $j \leqslant 1$; (ii) the sphericity and horizon linear velocity are bounded by the same existence line. Thus, both families can only have the same values of sphericity and horizon linear velocity as the Kerr ones, which implies that all BHs have a horizon which is  an oblate spheroid and the rotation of its null generators (relatively to a static observer at spatial infinity) never exceeds the speed of light; (iii)  the topology of the ergo-regions is either an ergo-sphere or an ergo-Saturn. In both families, the ergo-Saturn only appears for the solutions with the lower values of angular frequency.

In this work, we also presented a study of the structure of COs. Counter-rotating COs show a more complex structure than the co-rotating COs counterpart. Concerning the former, we saw that solutions with low or moderate amounts of hair (\textit{e.g.} $q = 0.5$) will have a structure similar to the Kerr one. For very hairy solutions ($q \geqslant 0.8$), a new region of UCOs can emerge, as well as a new region of No COs when the axionic hair outside the event horizon is compact enough to yield an extra pair of LRs. In the extreme case, where most of the BH solution is composed of axionic hair ($q = 0.999$), it possible to have a third region of UCOs, leading to an intercalation of SCOs and UCOs regions. 

Let us briefly comment on energy conditions. In \cite{Delgado:2020udb}, it was shown that the weak energy condition (WEC) and the dominant energy condition (DEC) always hold for the case of axion boson stars, whereas the strong energy condition (SEC) could be violated, and in fact, it is violated for a zero angular momentum observer (ZAMO). For the BH generalisation we can prove, following~\cite{Delgado:2020udb}, that the WEC and DEC are never violated, but the SEC can be violated, for instance, for a ZAMO.

Due to the groundbreaking results presented by the LIGO-Virgo collaboration detecting several gravitational waves events, starting with~\cite{Abbott:2016blz}, as well as the results by the EHT collaboration on the first image of a shadow of the M87 supermassive BH~\cite{Akiyama:2019cqa}, two possible and interesting directions to follow up on this work are: (1) to study the possible gravitational waves generated by the collision of two axionic hairy BHs, both in the head-on scenario, as well as, in the more realistic scenario of an in-spiral binary system. Recently~\cite{CalderonBustillo:2020srq} such collisions were made for Proca stars~\cite{Brito:2015pxa} obtaining a suggestive agreement with the particular gravitational wave event GW190521~\cite{Abbott:2020tfl}; (2) and to study the shadow of axionic hairy BHs, to analyse the impact of the QCD axion-like potential in the BH shadows. This would be a generalisation of the analysis in~\cite{Cunha:2015yba}.

\section*{Acknowledgements}

J. D. is supported by the FCT grant SFRH/BD/130784/2017. This work is supported by the Center for Research and Development in Mathematics and Applications (CIDMA) through the Portuguese Foundation for Science and Technology (FCT - Funda\c c\~ao para a Ci\^encia e a Tecnologia), references UIDB/04106/2020 and UIDP/04106/2020 and by national funds (OE), through FCT, I.P., in the scope of the framework contract foreseen in the numbers 4, 5 and 6 of the article 23, of the Decree-Law 57/2016, of August 29, changed by Law 57/2017, of July 19. We acknowledge support from the projects PTDC/FIS-OUT/28407/2017,  CERN/FIS-PAR/0027/2019 and PTDC/FIS-AST/3041/2020. This work has further been supported by the European Union's Horizon 2020 research and innovation (RISE) programme H2020-MSCA-RISE-2017 Grant No.~FunFiCO-777740. The authors would like to acknowledge networking support by the COST Action CA16104.

\bibliography{References}
\bibliographystyle{ieeetr}

\end{document}